\documentclass[final,5p,times,twocolumn]{elsarticle}
\usepackage{graphicx}
\usepackage{textcomp}
\usepackage{amsmath,amssymb,amsfonts}
\usepackage{algorithmic}
\usepackage{graphicx}
\usepackage{caption}
\usepackage{textcomp}
\usepackage{color}
\usepackage[dvipsnames]{xcolor}
\usepackage{subfigure}
\usepackage{multirow}
\usepackage{rotating}
\usepackage{diagbox}
\usepackage{stfloats}
\usepackage[backref]{hyperref}
\usepackage{float}
\usepackage{bm}
\usepackage{booktabs}
\usepackage{float}
\usepackage{multirow}
\usepackage{longtable}
\usepackage{lscape}
\usepackage{array}

\usepackage{amssymb}

\journal{Elsevier Journal}

\begin{document}

\begin{frontmatter}



\title{
A State-of-the-art Survey of Artificial Neural Networks for Whole-slide Image Analysis: 
from Popular Convolutional Neural Networks to Potential Visual Transformers}

\author{
Xintong Li$^1$, Weiming Hu$^1$, Chen Li$^1$, Tao Jiang$^2$, 
Hongzan Sun$^3$, Xiaoyan Li$^3$, Xinyu Huang$^4$, Marcin Grzegorzek$^4$ \\
1. Microscopic Image and Medical Image Analysis Group, 
College of Medicine and Biological Information Engineering, 
Northeastern University, Shenyang, China \\
2. Chengdu University of Information Technology, Chengdu, CHina
3. China Medical University, Shenyang, China \\ 
4. Institute for Medical Informatics, University of Luebeck, Luebeck, Germany \\
Corresponding author: Chen Li, E-mail: lichen201096@hotmail.com}

\begin{abstract}
In recent years, with the advancement of computer-aided diagnosis (CAD) technology and whole slide image (WSI), histopathological WSI has gradually played a crucial aspect in the diagnosis and analysis of diseases. To increase the objectivity and accuracy of pathologists' work, artificial neural network (ANN) methods have been generally needed in the segmentation, classification, and detection of histopathological WSI. In this paper, WSI analysis methods based on ANN are reviewed. Firstly, the development status of WSI and ANN methods is introduced. Secondly, we summarize the common ANN methods. Next, we discuss publicly available WSI datasets and evaluation metrics. These ANN architectures for WSI processing are divided into classical neural networks and deep neural networks (DNNs) and then analyzed. Finally, the application prospect of the analytical method in this field is discussed. The important potential method is Visual Transformers.
\end{abstract}


\end{frontmatter}


\section{Introduction}
\label{Intro}
\subsection{Brief Knowledge}
In pathology, biopsy diagnosis is the gold standard for cancer diagnosis~\cite{Kumar-2005-RCPB}. Conventionally, the inspection of slides under a microscope has been a typical method in histopathology~\cite{Della-2006-UAAD, Evered-2011-APVM}, and over the past decade, automated analysis of histopathology has become one of the fastest-growing areas in medical image calculating. \emph{Computer-aided Diagnosis} (CAD) can reduce the workload of pathologists and the rate of misdiagnosis~\cite{Lin-2019-FSFD}.
The digital version to slide is called \emph{Whole-slide Image} (WSI). It allows pathology slides to be stored digitally on a computer.

\begin{figure}[htbp!]
\centerline{\includegraphics[width=0.98\linewidth]{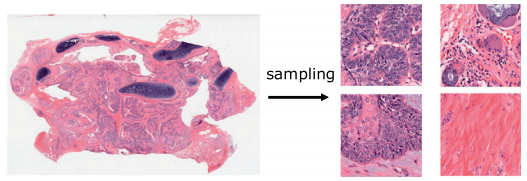}}
\caption{The process of cutting WSI into a patch. The figure matches to Fig.1 in~\cite{Komura-2018-MLMH}.}
\label{fig:2018-EMLMH}
\end{figure}

WSI, also known as virtual microscopy, involves digitizing or scanning slides to make digital slides for human observation and automatic image analysis. Currently, WSI technology has advanced to the point where digital slide scanners can make high-resolution digital images in a relatively short time. The digital slide show is designed to simulate a light microscope. These WSIs can be checked on the computer screen with the assistance of persons working in the relevant field. The viewer can examine the whole slide in a manner comparable to that of a light microscope and observe it in any direction and at different magnification~\cite{Glatz-2006-FKMW, Schrader-2006-DPUV, Romer-2003-UMSM}. However, due to the large resolution level and pixel size of WSI, this poses a huge challenge for image-level processing. If the thumbnails are down-sampled, a lot of information will be lost. Therefore, the usual processing method is to intercept WSI into patches with fine details for analysis~\cite{Wang-2018-WSLW}. The process of this interception is shown in the Figure.~\ref{fig:2018-EMLMH}.

In recent years, WSI has provided convenience in various fields, especially in the field of pathological research. For example, many people can open the same WSI at different places at the same time. This feature helps pathologists remote viewing, conferences, and E-learning. And its high quality, low storage cost, non-destructive, and fading characteristics are conducive to preservation. In the part of patients, the nursing of patients is further improved. However, the regulatory issues and practice standards of WSI are also issues that need attention and consideration~\cite{Pantanowitz-2011-RCSW}.

The use of CAD for WSI viewing is an efficient, accurate, and intuitive method. It can be used to diagnose different diseases, such as lung cancer, gastric cancer, colon cancer, and breast cancer. For example, in the work of~\cite{Wang-2019-WSDL}, a fasting and efficient classification method for WSI lung cancer images is proposed. Firstly, the patch-based \emph{full convolutional network} (FCN) is used to retrieve images at the patch-level, which is used to extract depth features and generate WSI descriptors. Then the descriptor is input into the Random Forest (RF)~\cite{Cutler-2001-PPRT} classifier used for image-level prediction to classify tumors. Finally, the high accuracy rate of this method is 97.3\%. The study of~\cite{Sharma-2017-DCNN}, proposed a \emph{Convolutional Neural Network} (CNN)~\cite{Bouvrie-2006-NCNN} architecture for WSI classification and necrosis detection of gastric cancer histopathology. This method is compared with the use of traditional features combined with RF for classification and AlexNet~\cite{Krizhevsky-2012-ICDC} for comparative analysis. According to the final experiment, the proposed CNN architecture has obtained favorable results. The overall classification accuracy of cancer classification is $69.90\%$, and the overall classification accuracy of necrosis detection is $81.44\%$. In the domain of colon cancer, the aim of~\cite{Jiao-2013-CCDW} propose the classification method derived from Support Vector Machine (SVM)~\cite{Cortes-1995-SVM} to classify colon cancer and non-colon cancer WSI. Then 18 simple features such as gray-level mean, gray-level variance, etc., and 16 texture features are selected. These 16 features are extracted by the Gray-level Co-occurrence Matrix (GLCM). Then the extracted features are put together as feature sets. The final experimental results are that the mean values of accuracy, recall, and F-score are $96.67\%$, $83.33\%$, and $89.51\%$, respectively. The experimental results show the important contribution of this method in classifying colon cancer WSI. The author of~\cite{Bejnordi-2016-ADDW} proposed a multi-scale superpixel classification method to detect the epithelial area of Ductal Carcinoma In Situ (DCIS) WSI, then classify it and detect the scope of DCIS. The final test results detected abnormal WSI in $80.00\%$ and $83.00\%$ of DCIS lesions, verifying the effectiveness of the method.

\subsection{The Development of WSI Analysis Using ANN}
\begin{figure}[htbp!]
\centerline{\includegraphics[width=0.98\linewidth]{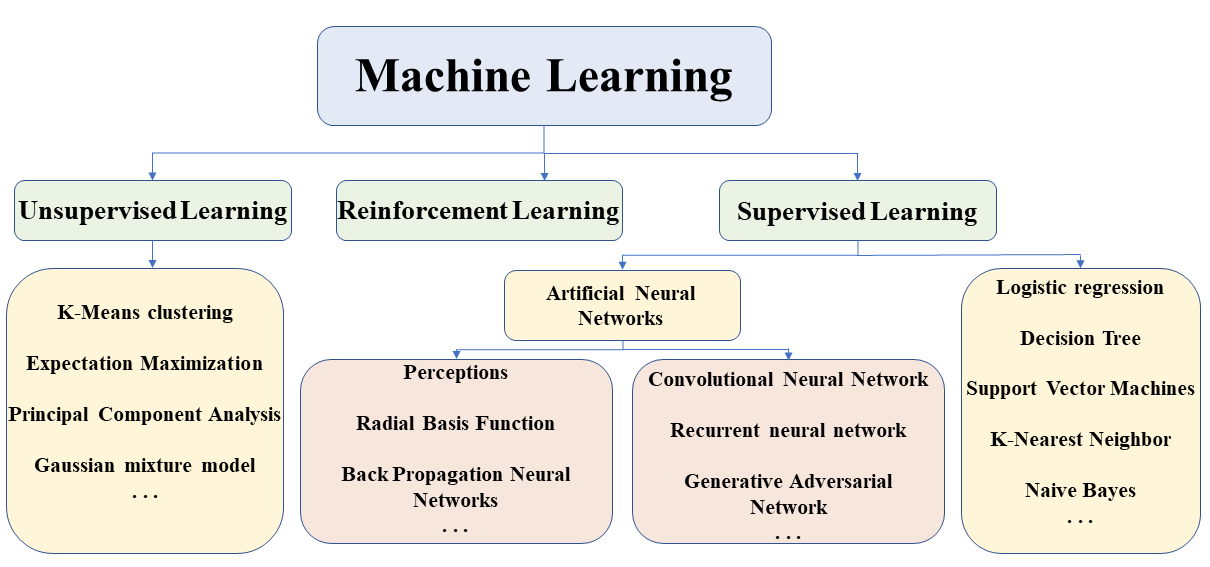}}
\caption{The structure of ANN technology in the ML system.}
\label{fig:MLintroductioon}
\end{figure}

Based on the rapid development of these CAD methods, more and more scholars begin to enter the field of \emph{Machine Learning} (ML). ML is a special research of how computers simulate or realize human learning behavior, to achieve modern knowledge or skills, reconstruct the current knowledge framework, and constantly increase their achievement. ML can separate into supervised learning, unsupervised learning, and reinforcement learning according to the types of data processed~\cite{Mohri-2018-FML, Bishop-2006-PRML}. As shown in Figure.~\ref{fig:MLintroductioon}.

\begin{figure}[htbp!]
\centerline{\includegraphics[width=0.6\linewidth]{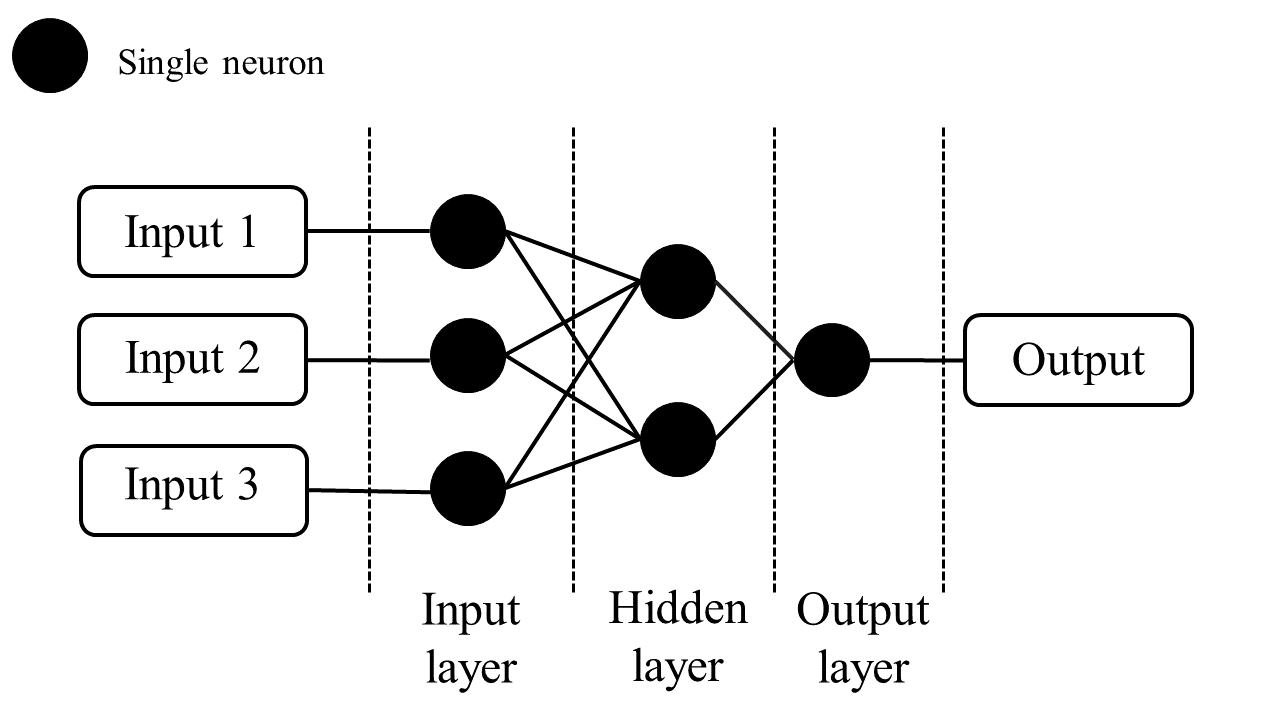}}
\caption{The most basic neural network model.}
\label{fig:2011-ANNM}
\end{figure}

Among different machine learning methods, \emph{Artificial Neural Network} (ANN)~\cite{Mcculloch-1943-LCII} is an important technology to simulate the intelligent principle of the human brain. The meaning of ANNs is fundamentally proposed from the biological field. In this field, neural networks play a critical part in the human body. The human body does what it should do to assist neural networks. A neural network is a network that millions of linked neurons working together. With the help of these neurons, full coordinate processing is done in the human body, which is the best example of parallel processing~\cite{Sharma-2012-CSAN}.   Similar to biological neurons, ANNs have artificial neurons that also receive inputs from other components or other artificial neurons, and then convert the results into outputs via transfer functions after the inputs are weighted and added~\cite{Dharwal-2016-AANN}. The most basic neural network model is shown in Figure.~\ref{fig:2011-ANNM}, which usually consists of three parts: The input layer, the hidden layer, and the output layer. When ANN is operated on the computer, it is not used for specific functions to program. Instead of this function, they are trained as a whole in the dataset until they understand the data pattern presented to them. After training them, new patterns can be presented to them for functions such as classification, segmentation, and detection~\cite{Kalogirou-2001-ANNR}.

\begin{figure}[htbp!]
\centerline{\includegraphics[width=0.98\linewidth]{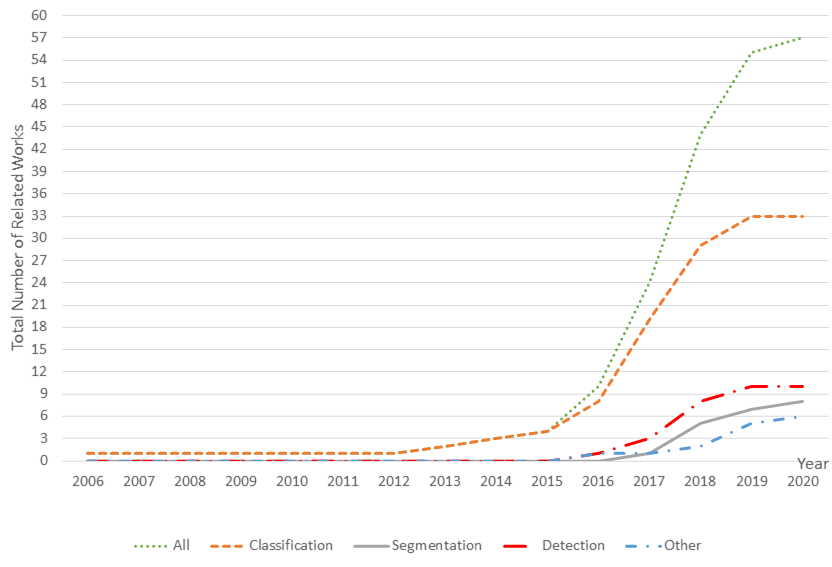}}
\caption{The pathologist is assisted by ANN to view the development trend of WSI.}
\label{fig:ANN-trend}
\end{figure}

Currently, ANNs are used to assist pathologists to browse the development trend of WSI as shown in Figure.~\ref{fig:ANN-trend}. Since 2006, ANNs have been used to assist pathologists in WSI. The three most commonly used fields are image classification, segmentation, and detection. From 2006 till now, the number of cases in these three main applications has increased year by year. Among them, the cases applied to classification are the most, followed by detection, then segmentation, and other applications such as machine-generated text~\cite{Maksoud-2019-CCOR} and automatic labeling~\cite{Sheikhzadeh-2016-ALMB}. The growth rate of its number is accelerating year by year, which also shows the progress of technology.

\subsection{Motivation of This Review}
As far as we know, there are some literature reviews on the related studies of ANNs that assist pathologists to view the work of WSI (e.g., the reviews
in~\cite{Zhou-2020-CRBH, Alzubaidi-2017-CADD, Goacher-2017-DCWS, farahani-2015-WSIP}). In the following part, we go through the survey papers.

The survey of~\cite{Zhou-2020-CRBH} reviews the methods of breast cancer histopathological image analysis based on ANNs, including segmentation, classification, and dataset partition. More than 130 papers are summarized, but they are only about breast cancer, not about other kinds of cancer.

The survey of~\cite{Alzubaidi-2017-CADD} summarizes the important algorithms of CAD of lung tissue, including traditional deep learning algorithms and ANNs. More than 20 papers are summarized, but in this paper, we only focus on the tasks based on ANNs and WSI technology.

In~\cite{Goacher-2017-DCWS}, the comparison of pathological diagnosis results by WSI and light microscopy are reviewed to measure the consistency of the two methods. More than 30 studies have been summarized, but the deep learning algorithms of CAD are not analyzed.

The survey of~\cite{farahani-2015-WSIP} summarizes the application of WSI technology platforms. According to clinical and non-clinical classification. More than 60 papers are summarized, but they are all about the hardware of WSI systems and the advantages, limitations, and prospects of their development.

\begin{figure}[htbp!]
\centerline{\includegraphics[width=0.7\linewidth]{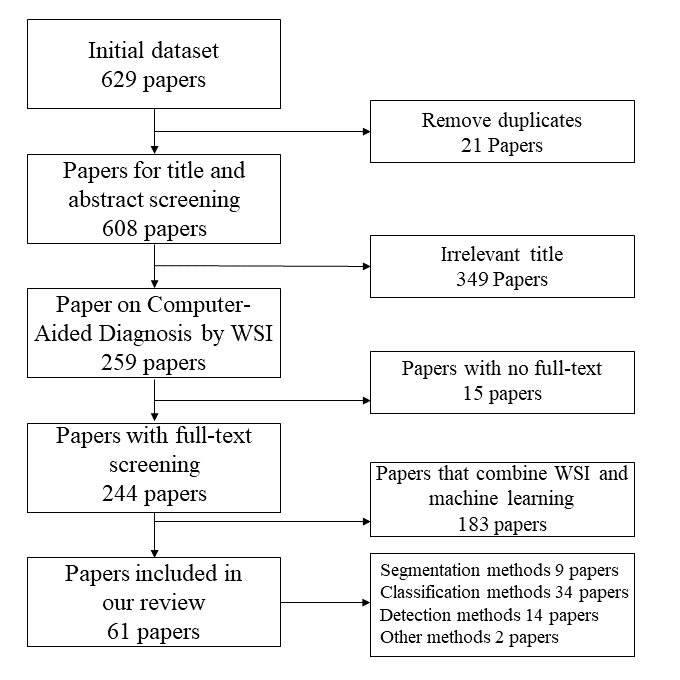}}
\caption{The systematic flow chart of paper selection for our work.}
\label{fig:paper selection}
\end{figure}

From the current review papers, we can find that many researchers are concerned about the device hardware, development status, and trend of WSI technology. In these review papers, many related works are methodically summarized and discussed. However, all of these research papers are not intended to introduce the use of ANN computer-aided pathologists to review WSI technology in detail. Therefore, we propose this review paper, combined with the related work done by ANN and WSI in the past few decades, using CAD to analyze. This survey summarizes more than 50 related works from 2006 to 2020. The process of selecting papers is shown in Figure.~\ref{fig:paper selection}. The audience of this review is related researchers and medical professionals in the field of medical imaging.
\subsection{Structure of This Review}
This structure of the paper is as follows: Sect. \ref{Common} illustrates the common ANNs methods. 
Sect. \ref{s:DEM} summarizes the related datasets and commonly used evaluation methods. 
Sect. \ref{TNNM}, and \ref{DNNM} present related work of ANN using WSI and CAD technology. 
After the overview of different works, the most frequently used approaches are analyzed
in Sect. \ref{MA}. Finally, Sect. \ref{PM} and \ref{s:CFW} conclude this review with prospective future direction.
\section{Common ANNs methods}
\label{Common}
In the previous description, we found that the CNN model is the most commonly used supervised machine learning model for WSI of CAD. Its inspiration comes from the tissue of the animal visual cortex~\cite{Hubel-1968-RFFA, Fukushima-1982-NSON}, it is designed to automatically and adaptively learn the spatial levels of features from low-level to high-level patterns. In 1980, a new neuron was considered to be the predecessor of CNN in~\cite{Fukushima-1983-NNNM}. LeNet was the pioneering work of LeCun et al. in CNNs in 1990~\cite{Le-1989-HDRB}, which was later improved. So far, CNN has been more widely used.

CNN, as the name implies. It uses a mathematical linear operation called convolution. This kind of linear operation can replace the original matrix multiplication operation in at least one layer of CNN~\cite{Dhillon-2020-CNNR}. It generally includes a convolutional layer, pooling layer, and fully connected layer~\cite{Shea-2015-ICNN}. The convolutional layer is the most important one. The convolution layer is mainly used to extract the input data. Here, the input data contains the features of multiple convolution cores. Each element of the convolution kernel corresponds to its weight co-efficient and deviation. It's like neurons in a feedforward neural network. Each neuron in the area of the current convolution layer near the previous convolution layer is connected to several neurons. The size of the convolution layer determines the size of this area, namely the receptive field, which can be compared with the receptive field of visual cortex cells~\cite{Gu-2018-RACN}. When the convolution kernel works, it will scan the input features in a fixed time. The operation also includes multiplication and summation of the matrix elements generated by the input features in the receiving domain and then stacking the deviations~\cite{Goodfellow-2016-DL}.

There are two kinds of pooling layers, which are maximum pooling and average pooling. Its main function is to reduce the dimension, and reduce the dimension by downsampling the convolutional results~\cite{Dhillon-2020-CNNR}. The full convolutional layer is to expand the output of the convolutional layer and pooling layer into a one-dimensional form, followed by a regression network or classification network with the same structure as the ordinary network, which is generally connected behind the pooling layer~\cite{Yamashita-2018-CNNO}. 

\begin{figure}[htbp!]
\centerline{\includegraphics[width=0.98\linewidth]{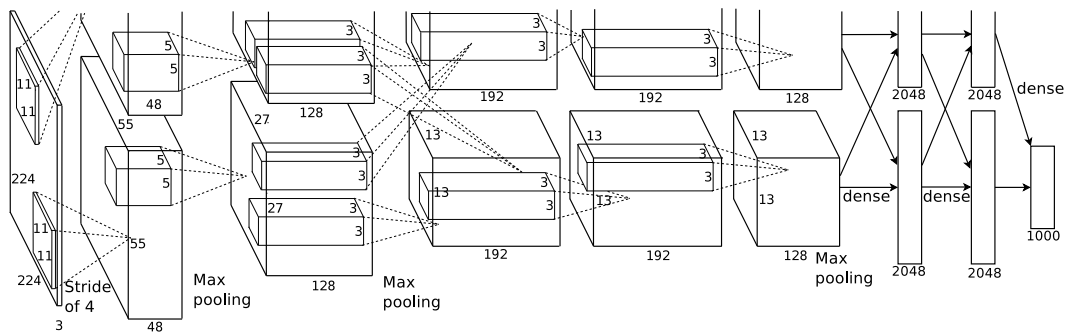}}
\caption{The network structure diagram of LeNet. The figure matches to Fig.2 in~\cite{Nayak-2013-CTHS}.}
\label{fig:LeNet}
\end{figure}

LeNet is the originator of CNN network structure, which defines CNN network architecture for the first time. In 2012, Krizhevsky and Hinton launched AlexNet~\cite{Krizhevsky-2012-ICDC}. For the first time, AlexNet applied ReLu and dropout in CNN to avoid overfitting the model, and its network architecture is deeper than that of LeNet. In the ILSVRC (Imagenet Large-scale Visual Recognition Challenge) of that year, the accuracy rate of AlexNet exceeded second place by 10.9\%, which caused many scholars to study deep learning and officially opened the prelude of deep learning. The network structure diagram of LeNet is shown in Figure.~\ref{fig:LeNet}.

In 2014, Simonyan and Zisserman proposed the VGG series model, which appeared on ILSVRC2014 as the second-ranked basic network for classification tasks and the first for positioning tasks. For the time, VGG belonged to a deeper network, and it can be seen as a deepened version of AlexNet. VGG is composed of 5 layers of convolutional layers, 3 layers of fully convolutional layers, and a softmax output layer. The layers are separated by max-pooling, and the activation units of all hidden layers use the ReLu function. The pooling core is smaller, the number of channels is more, and the number of layers is deeper~\cite{Simonyan-2014-VDCN}. 

\begin{figure}[htbp!]
\centerline{\includegraphics[width=0.7\linewidth]{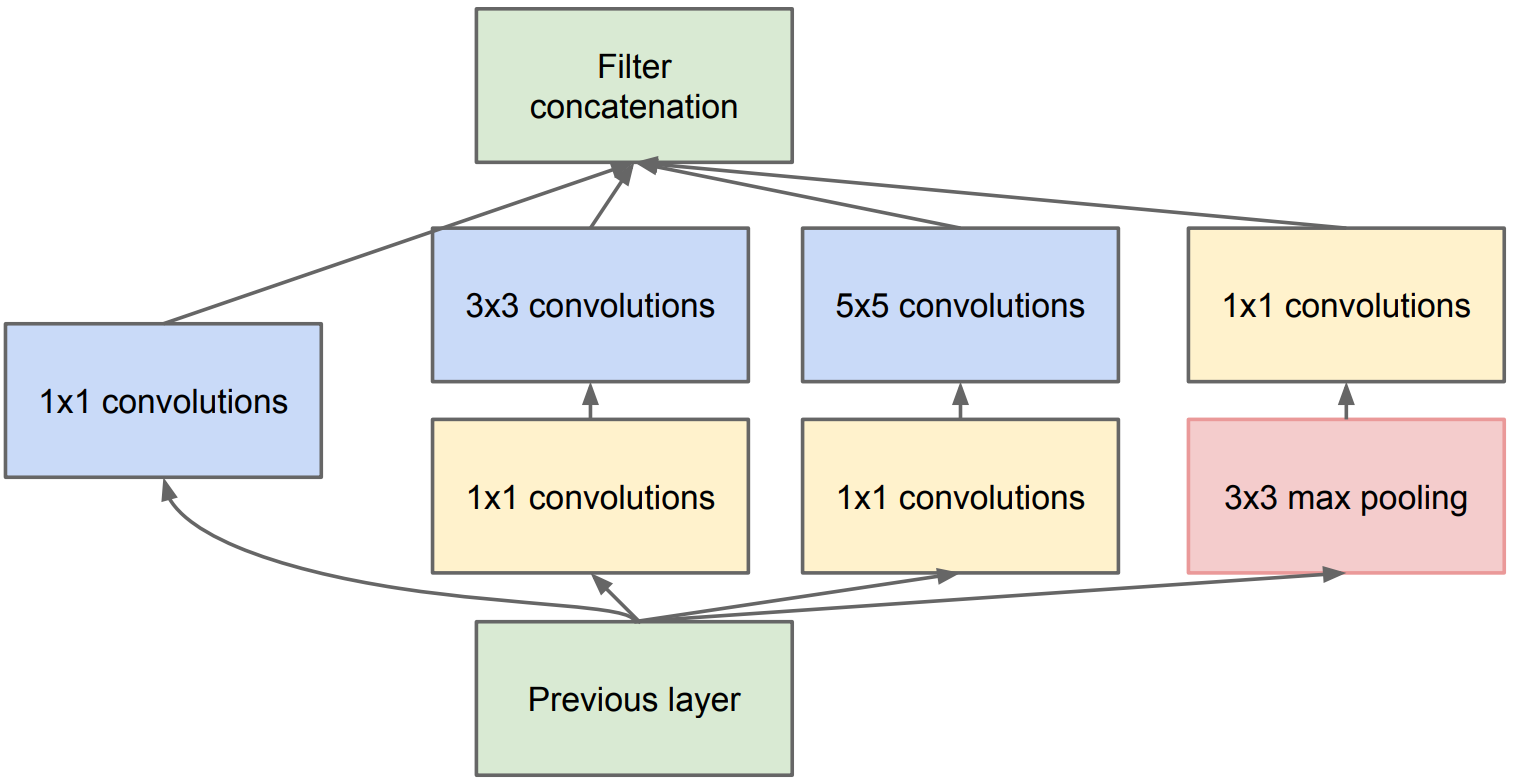}}
\caption{The Inception module with dimensionality reduction. The figure matches Fig.2 in~\cite{Szegedy-2015-GDC}.}
\label{fig:Inception-Net-V1}
\end{figure}

In the same year, Google proposed the Inception network structure and built GoogLeNet (also known as Inception Net V1)~\cite{Szegedy-2015-GDC} based on this. This architecture won first place in the classification and detection task of ILSVRC2014. Unlike VGG, Inception's structure is deeper and wider. The main innovation is to choose to connect the convolutional layer with the convolution kernel size of $1×1$, $3×3$, $5×5$ and the pooling layer with the pooling kernel size of 3×3 in parallel, and perform the feature maps obtained by each concatenate operations are merged as subsequent input. This method greatly improves the efficiency of classification. Next, in 2015, Google proposed the Batch Normalization operation, added it to GoogLeNet and modified a certain structure, and got Inception Net V2~\cite{Ioffe-2015-BNAD}. Then, Inception Net V3~\cite{Szegedy-2016-RIAC} and Inception Net V4~\cite{Szegedy-2016-IRIR} are successively proposed. The Inception module with dimensionality reduction is shown in Figure.~\ref{fig:Inception-Net-V1}.

\begin{figure}[htbp!]
\centerline{\includegraphics[width=0.6\linewidth]{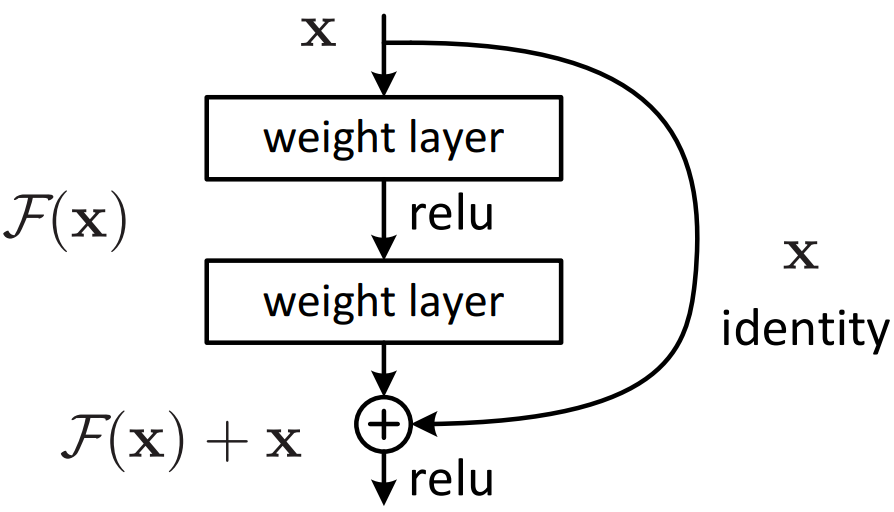}}
\caption{Residual learning: a building block. This figure matches Fig.2 in~\cite{He-2016-DRLI}.}
\label{fig:residual structure}
\end{figure}

\begin{figure}[htbp!]
\centerline{\includegraphics[width=0.7\linewidth]{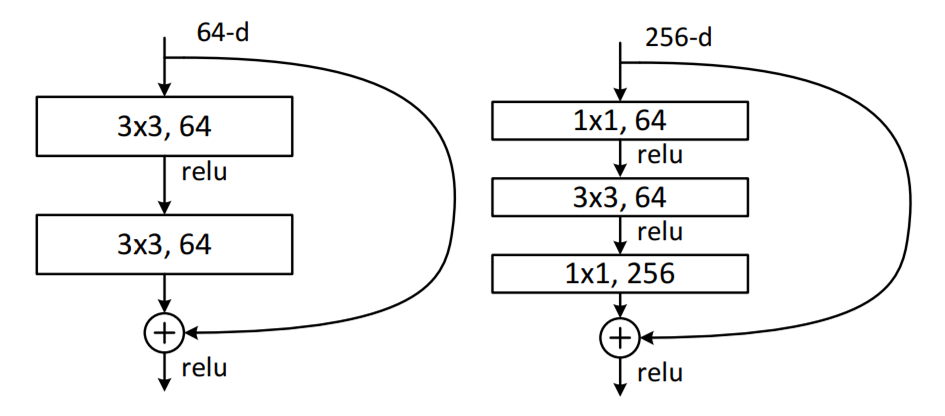}}
\caption{Left: a building block.  Right: a bottleneck building block for ResNet-50/101/152. This figure corresponds to Fig.5 in~\cite{He-2016-DRLI}.}
\label{fig:Turn}
\end{figure}

In 2015, Kaiming He proposed ResNet. The Residual Block in ResNet solves the problem of poor fitting ability of deep neural networks (DNNs) after reaching a certain depth and also alleviates the problem of gradient disappearance. The Figure.~\ref{fig:residual structure} is the residual structure. The original function $H(X)$ to be fitted is changed to $F(X)$, where $H(x) =F(x)+X$.  If the depth of the model continues to increase, as shown in Figure.~\ref{fig:Turn}, the original residual block is improved to bottleneck block to reduce the parameters and calculation of the model~\cite{He-2016-DRLI}.

\section{Datasets and Evaluation Methods}
\label{s:DEM} 
In this section, we have discussed some commonly used datasets and evaluation metrics for the 
classification, segmentation, and detection tasks.

\subsection{Publicly Available Datasets about WSI}
\label{ss:int:data}
To make better use of WSI technology to analyze CAD, we summarize some common open datasets in our research. Tab.~\ref{DB} shows the basic information for these datasets. The two most commonly used datasets are The Cancer Genome Atlas (TCGA)~\cite{TCGA} and the Camelyon datasets~\cite{Litjens-2018-1HSS}. These two datasets are usually used for classification and detection. Meanwhile, we notice two WSI datasets named TUPAC16~\cite{Veta-2019-PBTP} 
and Kimia Path24~\cite{Babaie-2017-CRDP}. TUPAC dataset is generally used, like mitosis detection, breast tumor proliferation prediction, automatic scoring (classification), etc. Kimia Path24 is regularly used for classification and retrieval~\cite{Babaie-2017-CRDP}~\cite{Kumar-2018-DBFR}. The basic information 
of the common datasets is shown in Table.~\ref{DB}.
\begin{table}[htbp!]\tiny
\renewcommand\arraystretch{2}
\setlength{\tabcolsep}{1mm}
\centering
\scriptsize\caption{The basic information of the publicly available used datasets.}
\begin{tabular}{clcc}
\hline
\textbf{Databases}    & \multicolumn{1}{c}{\textbf{Year}} & \textbf{Field}   & \textbf{Number of images or or size} \\ \hline
TCGA                  & 2006                              & Cancer related   & Over 470 TB                          \\
NLST Pathology Images & 2009                              & Lung             & Around 1250 H\&E slides              \\
BreakHis              & 2015                              & Breast cancer    & 9,109 microscopic images             \\
TUPAC16               & 2016                              & Tumor mitosis    & Around 821 H\&E slides               \\
Camelyon              & 2017                              & Breast cancer    & Around 3TB                           \\
Kimia Path24          & 2017                              & Pathology Images & 24 WSIs                              \\ \hline
\end{tabular}
\label{DB}
\end{table}

\subsubsection{TCGA database} 
\begin{figure}[htbp!]
\centerline{\includegraphics[width=0.8\linewidth]{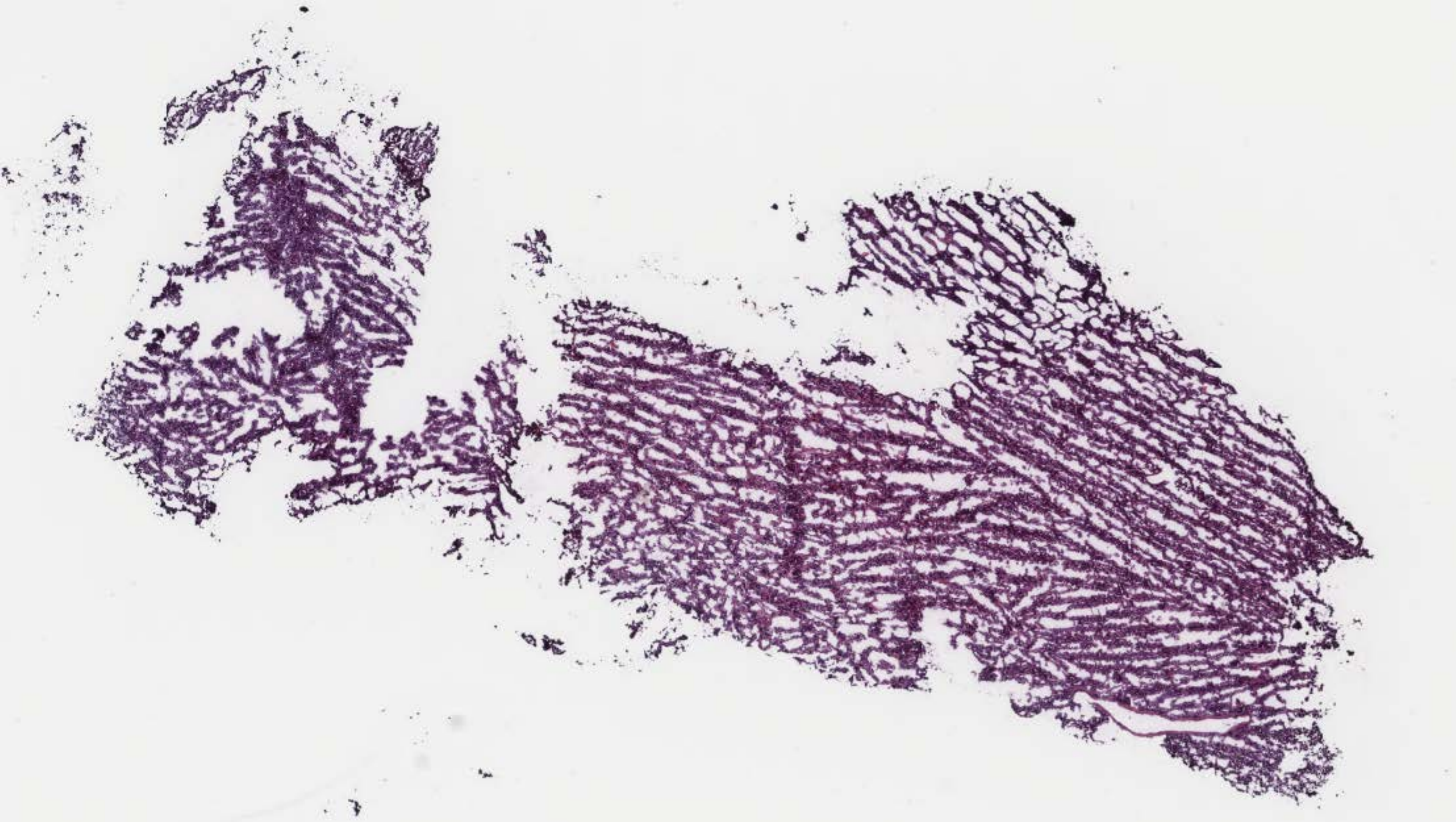}}
\caption{An example of WSI in an adrenal cortical carcinoma in the TCGA database~\cite{TCGA}.}
\label{fig:fig4}
\end{figure}

TCGA is a project jointly launched by the National Cancer Institute (NCI) and the 
National Human Genome Research Institute (NHGRI) in 2006~\cite{TCGA}. It contains 
clinical data, genome variation, mRNA expression, miRNA expression, methylation, and 
other data of various human cancers (including subtypes of tumors). The database is designed to apply high-throughput genomic analysis approaches to assist people in developing a better considerate of cancer and improve the ability to prevent, diagnose, and treat~\cite{Tomczak-2015-CGAI}. 
While TCGA fundamental task concentrates on genomics and clinical data, it also acquires a large number of WSIs in patient’s tissue. Since WSI datasets are much larger than other datasets, to speed viewing, David et al.~\cite{Gutman-2013-CDSA} propose a combined network platform named Cancer Digital Slide Archive (CDSA) to contain all WSIs in TCGA. Because the dataset contains many types of cancers, it has a deep range of applications. 
Fig.~\ref{fig:fig4} is an instance of WSIs in an adrenal cortical carcinoma in the TCGA database.

\subsubsection{Camelyon Database} 
The Camelyon Challenge is hosted by International Symposium on Biomedical Imaging (ISBI)~\cite{Litjens-2018-1HSS}. The whole competition dataset (Camelyon16, Camelyon17) are derived from sentinel lymph nodes of breast cancer patients contains WSIs of Hematoxylin and Eosin (H\&E) stained node sections~\cite{Bejnordi-2017-DADL}\cite{Bandi-2018-DIMC}. 

\begin{figure}[htbp!]
\centerline{\includegraphics[width=0.8\linewidth]{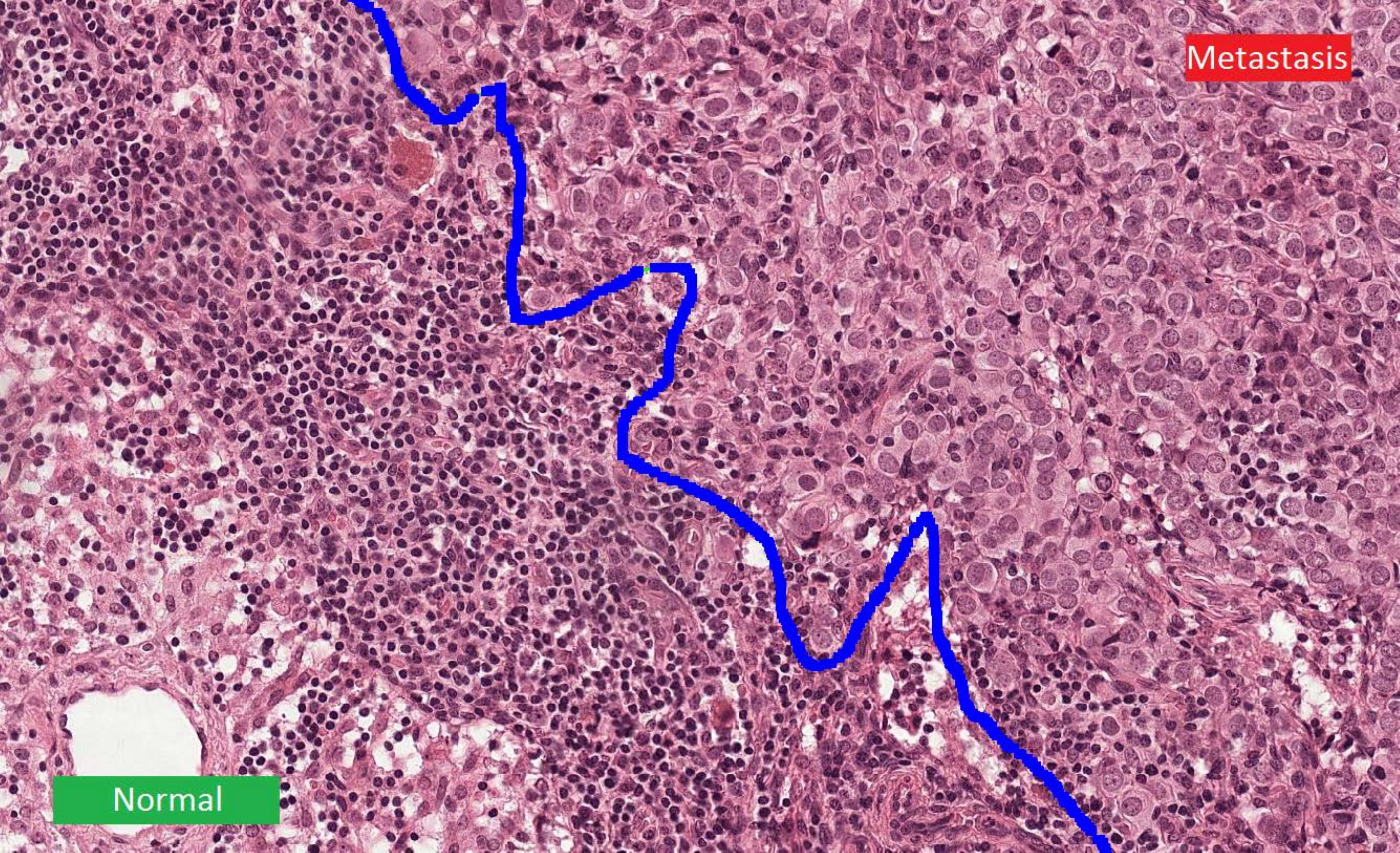}}
\caption{A pathological image of a lymph node in Camelyon~\cite{Litjens-2018-1HSS}. 
The left side belongs to normal cell tissue, 
and the right cell has been swallowed and occupied by cancer cells.}
\label{fig:fig5}
\end{figure}

Accordingly, the Camelyon dataset is fitting for the automatic detection and classification of breast cancer in WSI. The data of Camelyon16 are from the Radboud University Medical Centre (RUMC) and the University of Utrecht Medical Centre (UMCU). The Camelyon16 dataset is composed of 170 phase I lymph node WSIs (100 normals and 70 metastatic) and 100 Phase II WSIs (60 normals and 40 metastatic), and the test dataset includes 130 WSIs from two universities. The Camelyon16 dataset is used as training values for the evaluation of Camelyon17. 
Fig.~\ref{fig:fig5}  is a pathological image of a lymph node in Camelyon. The left side refers to normal cell tissue, and the right cell has been engulfed and held by cancer cells.

\subsubsection{TUPAC16 Database}
\begin{figure}[htbp!]
\centerline{\includegraphics[width=0.9\linewidth]{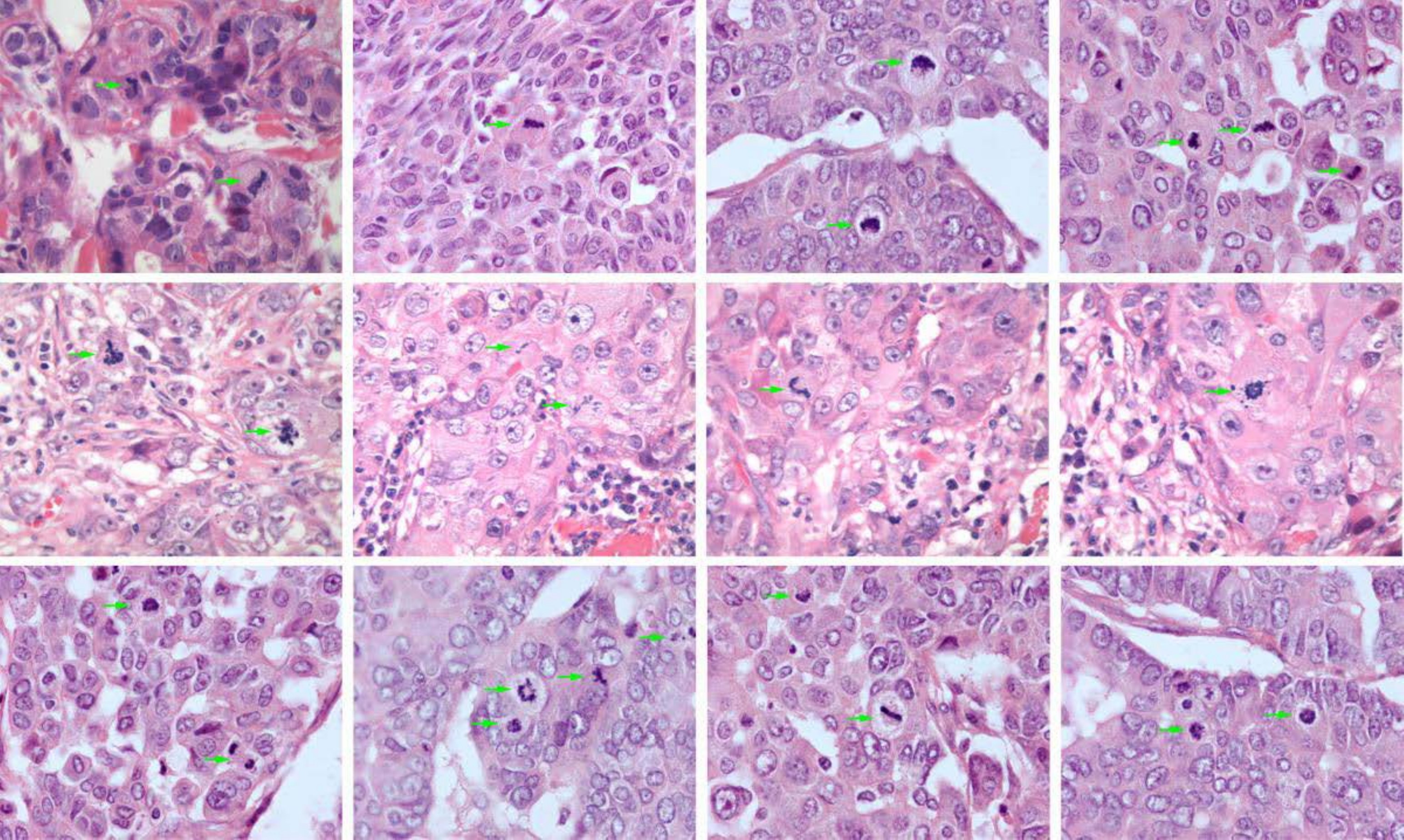}}
\caption{Some examples of mitosis diagrams in H\&E breast cancer slices in the 
TUPAC16~\cite{Veta-2019-PBTP}, with green arrows marking mitosis.}
\label{fig:fig6}
\end{figure}

The TUPAC16 challenge is held in the background of the MICCAI~\cite{Veta-2019-PBTP}. TUPAC16 main challenge dataset includes 821 TCGA WSIs with two kinds of tumor proliferation data. 500 for training, 321 for testing. In addition to the main challenge dataset, there are two extra datasets (area of interest and mitotic detection). 
The area of interest secondary dataset involves 148 cases that are inconsistently picked out from the training dataset. The mitotic test dataset involved WSIs of 73 breast cancer WSIs from three pathological centers. Of the 73 cases, 23 are AMIDA13 challenge~\cite{Veta-2015-AAMD}. The resting 50 cases previously used to assess the interobserver agreement for mitosis counting are from two other pathology centers in the Netherlands. Therefore, datasets are mainly used for automatic detection of mitosis or other regions of interest (ROI) in tumors. Fig.~\ref{fig:fig6} shows some examples of mitosis maps in H\&E breast cancer slices, with green arrows indicating mitosis.

\subsubsection{Kimia Path24 Database}
\begin{figure}[htbp!]
\centerline{\includegraphics[width=0.9\linewidth]{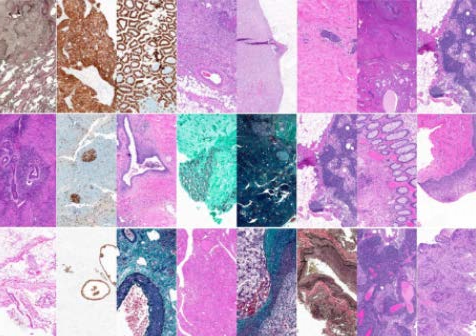}}
\caption{24 WSIs thumbnails in Kimia Path24 database~\cite{Babaie-2017-CRDP}.}
\label{fig:fig7}
\end{figure}

The dataset is artificially selected consciously. There are 350 WSIs from different body positions, so 24 WSIs represent different texture patterns. Therefore, this dataset is more like a computer vision dataset (opposite to the pathological dataset), because the visual attention is focused on the diversity of patterns, rather than on the medical field~\cite{Babaie-2017-CRDP}. Therefore, the dataset is mainly used for the classification and retrieval of histopathological images. The 24 WSIs miniatures in the dataset are shown in Fig.~\ref{fig:fig7}.

\subsection{Evaluation Method}
\label{ss:int:EM}
This subsection introduces the evaluation methods of classification, segmentation, 
and detection algorithms and related formulas.

\subsubsection{Basic Evaluation Indexs} 
\label{ss:int:BEI}
The confusion matrix is used to observe the performance of the model in each category, 
and the probability of each category can be calculated. The specific style of the 
confusion matrix is shown in Tab.~\ref{CM}. 
\begin{table}[htbp!]
\renewcommand\arraystretch{2}
\setlength{\tabcolsep}{5mm}
\centering
\scriptsize\caption{Confusion matrix of basic evaluation indexs.}
\begin{tabular}{ccc}
\hline
\multicolumn{1}{l}{\textbf{Data Class}} & \multicolumn{1}{l}{\textbf{Classified as Pos}} & \multicolumn{1}{l}{\textbf{Classified as Neg}} \\ \hline
Pos                                     & True Positive(TP)                              & False Negative(FN)                             \\ \hline
Neg                                     & False positive(FP)                             & True Negative(TN)                              \\ \hline
\end{tabular}
\label{CM}
\end{table}

According to the confusion matrix, the True Positive Rate (TPR) can be defined as $TP/(TP + FN)$, 
which represents the proportion of the actual positive instances in all positive instances of 
the positive class predicted by the classifier, the False Postive Rate (FPR) can be defined 
as $FP/(FP+TN)$, which represents the proportion of the actual negative instances in all negative 
instances of the positive class predicted by the classifier. It can be seen that the mathematical 
expressions of the following evaluation metrics are shown in Table.~\ref{EM}~\cite{Sokolova-2009-SAPM}.
\begin{table}[htbp!]
\renewcommand\arraystretch{2}
\setlength{\tabcolsep}{5mm}
\centering
\scriptsize\caption{Evaluation metrics. \emph{Acc}, \emph{P}, \emph{R}, \emph{Se}, \emph{Sp} and \emph{F1} denote accuracy, precision, recall, 
sensitivity, specificity and F1 score, respectively.}
\begin{tabular}{cc|cc}
\hline
Assessments & Formula   &  Assessments & Formula                          \\ \hline
\emph{Acc}    & $\frac{TP + TN}{TP + TN + FP + FN}$  &  \emph{Se}	& $\frac{TP}{TP + FN}$ \\ 
\emph{P}   & $\frac{TP}{TP + FP}$   &   \emph{Sp}	& $\frac{TN}{TN + FP}$           \\
\emph{R}      & $\frac{TP}{TP + FN}$   &   \emph{F1}	&$2×\frac{P×R}{P + R}$         \\ \hline
\end{tabular}
\label{EM}
\end{table}

\subsubsection{Evaluation of Segmentation Methods} 
Image segmentation~\cite{Haralick-1985-IST} is the segmentation of images with existing 
targets and precise boundaries. The commonly used indicators are accuracy, precision, 
recall, F-measure, sensitivity, and specificity. These metrics we have discussed in 
Sec.~\ref{ss:int:BEI} and their mathematical expressions are given Tab.~\ref{CM}. 
Dice co-efficient (D) and Jaccard index (J) are popular segmentation evaluation indexes in 
recent years. Dice co-efficient (D) represents the ratio of the area intersected by two 
individuals to the total area, that is, the similarity between the ground truth and the 
segmentation result graph. If the segmentation is perfect, the value is 1. Then, if S 
stands for the segmentation result graph and G stands for ground truth, the expression 
of the Dice co-efficient (D) is given in Eq.~(\ref{eq:Dice}).
\begin{equation}
D(S,G)=\frac{2|A \cap G|}{|A| + |G|}
\label{eq:Dice}
\end{equation}

Jaccard Index (J) represents the intersection ratio of two individuals, which is similar 
to Dice co-efficient. The formula is given in Eq.~(\ref{eq:Jaccard}).
\begin{equation}
J(S,G)=\frac{|A \cap G|}{|A \cup G|}
\label{eq:Jaccard}
\end{equation}

\subsubsection{Evaluation of Classification Methods} 
\label{ss:int:EMC}
Classification~\cite{Kamavisdar-2013SICA} is the operation of determining the properties 
of objects in the image. In the field of digital histopathology we studied, some are the 
classification of cancer~\cite{Petushi-2006-LSCH}, some are the operation of selecting 
ROI~\cite{Swiderska-2015-TMMH}, and some are the identification of cancer 
regions~\cite{Doyle-2010-BBMC}. The purpose of classification is achieved by the constructed 
classifier. The performance indicators used to evaluate these classifiers are critical to 
the final results. Accuracy is the most commonly used indicator to evaluate classifiers. 
Precision, recall, sensitivity, specificity, and F1 score are widely used to evaluate 
classifiers. Accuracy,precision, recall, F-measure, sensitivity, and specificity we have 
discussed in Sec.~\ref{ss:int:BEI}, and their mathematical expressions are given in Tab.~\ref{CM}. 
With the continuous improvement of classification requirements in practical applications, 
ROC (Receiver Operating Characteristic), AUC (Area Under ROC Curve), a non-traditional 
measurement standard, has emerged. ROC is a curve drawn on a two-dimensional plane with 
FPR as the abscissa and TPR as the ordinate. It can reflect the sensitivity and specificity 
of the continuous variables as a comprehensive indicator. It can also solve the problem of 
class imbalance in the actual dataset. AUC quantifies the area under the ROC curve into a 
numerical value to make the results more intuitive.

\subsubsection{Evaluation of Detection Methods} 
Detection~\cite{Pal-1993-RIST} is another common task in analyzing histopathological WSIs. 
Detection is not only to determine the attributes of the region identified in WSI but also 
to identify and obtain more detailed results. Because of the similarity between testing and 
classification, most of the evaluation indexes are the same as the classification, including 
accuracy, precision, recall, F-measure, sensitivity, and specificity that we have discussed in 
Sec.~\ref{ss:int:BEI}. However, in WSI detection, it is difficult to locate, determine and 
quantify multiple lesions. Therefore, FROC (Free Receiver Operating Characteristic 
Curve)~\cite{Egan-1961-OCSD} is proposed to evaluate the detection results. FROC curve is 
a small variation of the ROC curve. It is a curve drawn on a two-dimensional plane with FP 
as the horizontal coordinate and TPR as the vertical coordinate. This allows the detection 
of multiple lesion areas on a single WSI.

\subsection{Summary}
According to the review above, we can see that the commonly used public datasets are TCGA, 
TUPAC16, and Kimia Path24 for the classification, segmentation, and detection of histopathological 
images using the combination of WSI technology and CAD with a brief introduction. Also, the evaluation indicators of these three tasks. The 
basic commonly used evaluation indicators are accuracy, precision, recall, sensitivity, 
and specificity. In terms of classification, there are comprehensive indicators such as AUC. 
Dice co-efficient and Jaccard index in segmentation indicators have become popular in 
recent years, and FROC in detection indicators can be used for positioning, qualitative, and quantitative analysis of multiple lesions. 

\section{Traditional Neural Network  Method}
\label{TNNM}
As can be seen from Figure.~\ref{fig:MLintroductioon}, ANN is separated into two parts. It is a shallow traditional neural network and a DNN. In the content, we investigated, traditional shallow neural networks are separated into Perceptrons, Radial Basis Function (RBF), Back Propagation Neural Networks, etc. The DNN is generally separated into CNN, Recurrent Neural Network (RNN) and Generative Adversarial Network (GAN) and so on.

In the papers we summarized earlier in the year, there were cases where shallow traditional neural networks were applied to WSI for CAD.
In~\cite{Petushi-2006-LSCH}, the LNKNET software package is used to classify breast cancer WSI. LNKNET integrates neural networks, machine learning, classification and statistics, clustering, and feature selection algorithms into a modular package to achieve the purpose of classification.

\begin{figure}[htbp!]
\centerline{\includegraphics[width=0.8\linewidth]{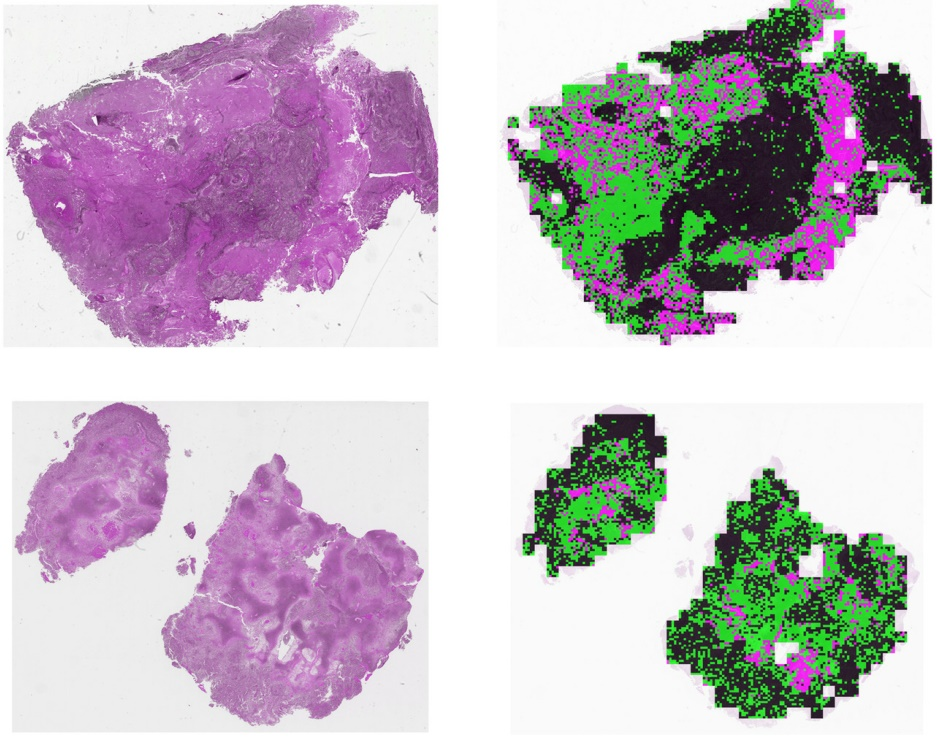}}
\caption{Two examples of classification results of a heterogeneous GBM tissue sections. The figure matches Fig.4 in~\cite{Nayak-2013-CTHS}.}
\label{fig:CTHS}
\end{figure}

In~\cite{Nayak-2013-CTHS}, an automatic feature learning method for WSI classification from unlabeled images is proposed. The first step is to learn the dictionary from unlabeled images. Then the sparse auto-encoder is used to learn the function. The automatic encoder consists of three parts: Encoder, dictionary, and a set of codes. The encoder is used to train the classifier on a small amount of label data. The multi-class regularization support vector classification is used. The regularization parameter is 1 and the polynomial kernel is 3. For data, two datasets from TCGA (i) glioblastoma multiforme (GBM) and (ii) clear cell renal cell carcinoma (KIRC) were used. The accuracy of classification is $84.3\%$ and $80.9\%$ respectively. The classification results of uneven GBM tissue sections are shown in the Figure.~\ref{fig:CTHS}.

\section{Deep Neural Network  Method}
\label{DNNM} 
\subsection{Segmentation Methods}
\label{s:SM} 

Among the papers we reviewed, 10 papers used deep learning methods for WSI 
segmentation~\cite{Bandi-2017-CDMT, Xu-2017-LSTH, Dong-2018-RANT, Cui-2018-DLAO, Sirinukunwattana-2018-IWSS, Mehta-2018-LSBB,  Seth-2019-ALBD, Seth-2019-ASDW, Feng-2020-DLMA}.

\begin{figure}[htbp!]
\centerline{\includegraphics[width=0.98\linewidth]{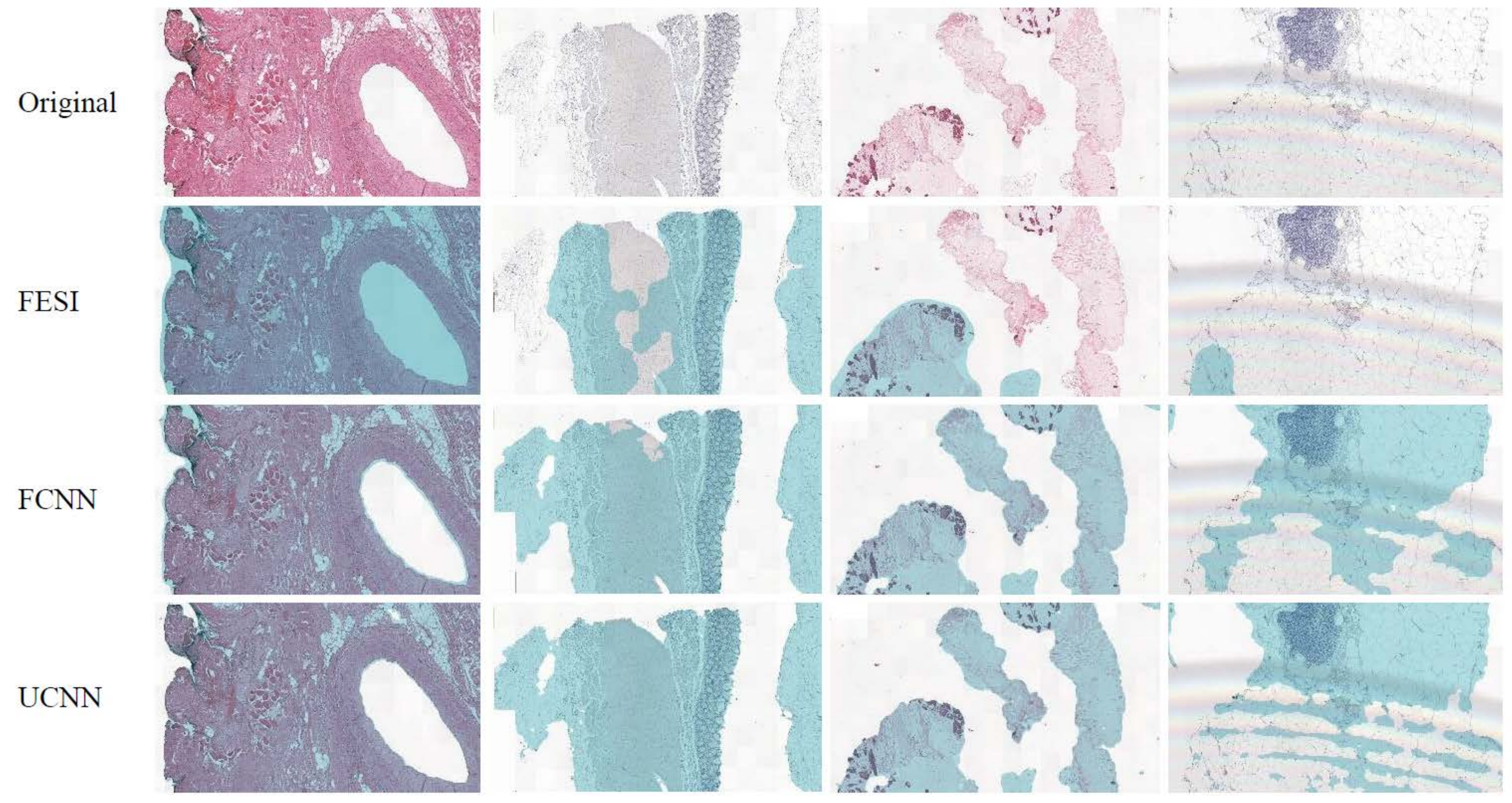}}
\caption{Qualitative results for the different algorithms. 
The figure matches to Fig.2 in~\cite{Bandi-2017-CDMT}}
\label{fig:CDMT}
\end{figure}

In~\cite{Bandi-2017-CDMT}, FCN and U-net are used to segment and accurately identify tissue slices. In this paper, the two methods are compared with the traditional foreground extraction (FESI) algorithm based on structural information. Finally, the three methods are applied to 54 WSIs and evaluated using the average value of the Yakoka index and its standard deviation. The final U-net results are the best (Jaccard index is $0.937$). Figure.~\ref{fig:CDMT} shows the qualitative effects of different algorithms.

In~\cite{Xu-2017-LSTH}, CNN-based Imagenet is used to extract features and transform the extracted features into histopathological images. Support vector machine (SVM) is used to define segmentation problems as classification problems. Finally, the method is applied to the digital pathology and colon cancer dataset of MICCAI 2014 challenge, and the final result win first place with $84\%$ accuracy.

In~\cite{Dong-2018-RANT}, an efficient and simple framework called Reinforced Auto-Zoom Net (RAZN) is proposed, which considers the accurate and fast prediction of breast cancer segmentation. RAZN learning strategy network to decide whether to enlarge a given ROI. Because the amplification is selective, RAZN is robust to unbalanced and noisy ground truth tags and effectively reduces overfitting. Finally, the method is evaluated on a public breast cancer dataset. It can be seen from the experimental results that RAZN is superior to single-scale and multi-scale baseline methods, and can obtain higher accuracy with lower cost.

In~\cite{Cui-2018-DLAO}, an automatic end-to-end DNN algorithm is proposed to segment the single core. The core boundary model is introduced to predict the core and its boundary simultaneously by FCN. Given the normalized color image, the model outputs the estimated kernel and boundary map directly. Then after the post-processing, the core of the final segmentation is produced. A method for extracting and assembling overlapping blocks is designed to predict the kernel in large WSI seamlessly. Finally, the results show that the data extension method is effective for the task of nuclear segmentation. The experiment shows that the method is superior to the existing technology and can segment WSI accurately in the acceptable time range.

\begin{figure}[htbp!]
\centerline{\includegraphics[width=0.98\linewidth]{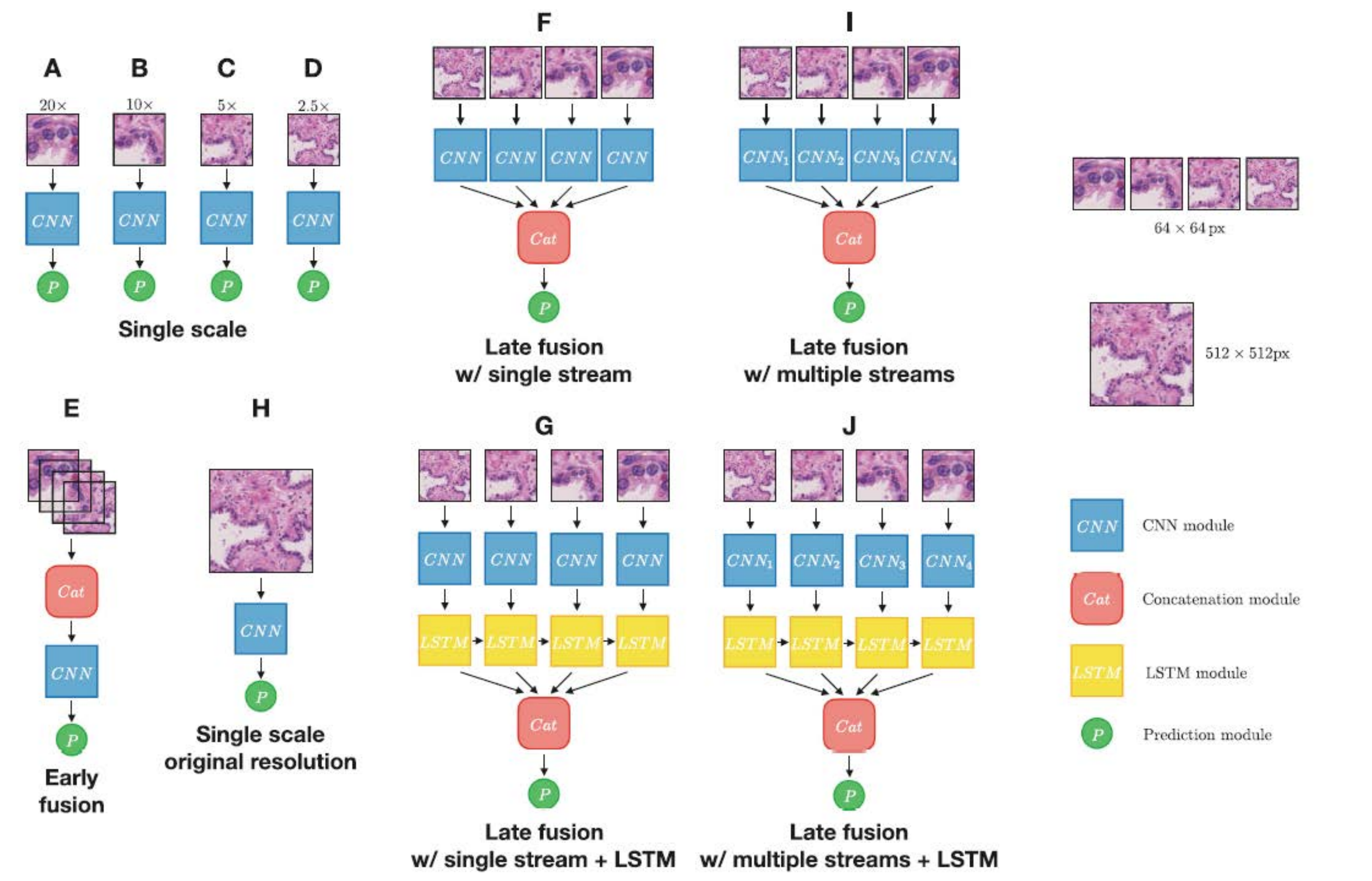}}
\caption{Used architectures. 
The figure matches to Fig.2 in~\cite{Sirinukunwattana-2018-IWSS}}
\label{fig:IWSS}
\end{figure}

In~\cite{Sirinukunwattana-2018-IWSS}, different architectures are systematically compared to evaluate how multi-scale information can affect segmentation performance. The architecture is shown in Figure.~\ref{fig:IWSS}. The experiment used a public breast cancer dataset and a locally collected prostate cancer dataset. The final results show that visual environment and scale play an important role in histological image classification.

\begin{figure}[htbp!]
\centerline{\includegraphics[width=0.98\linewidth]{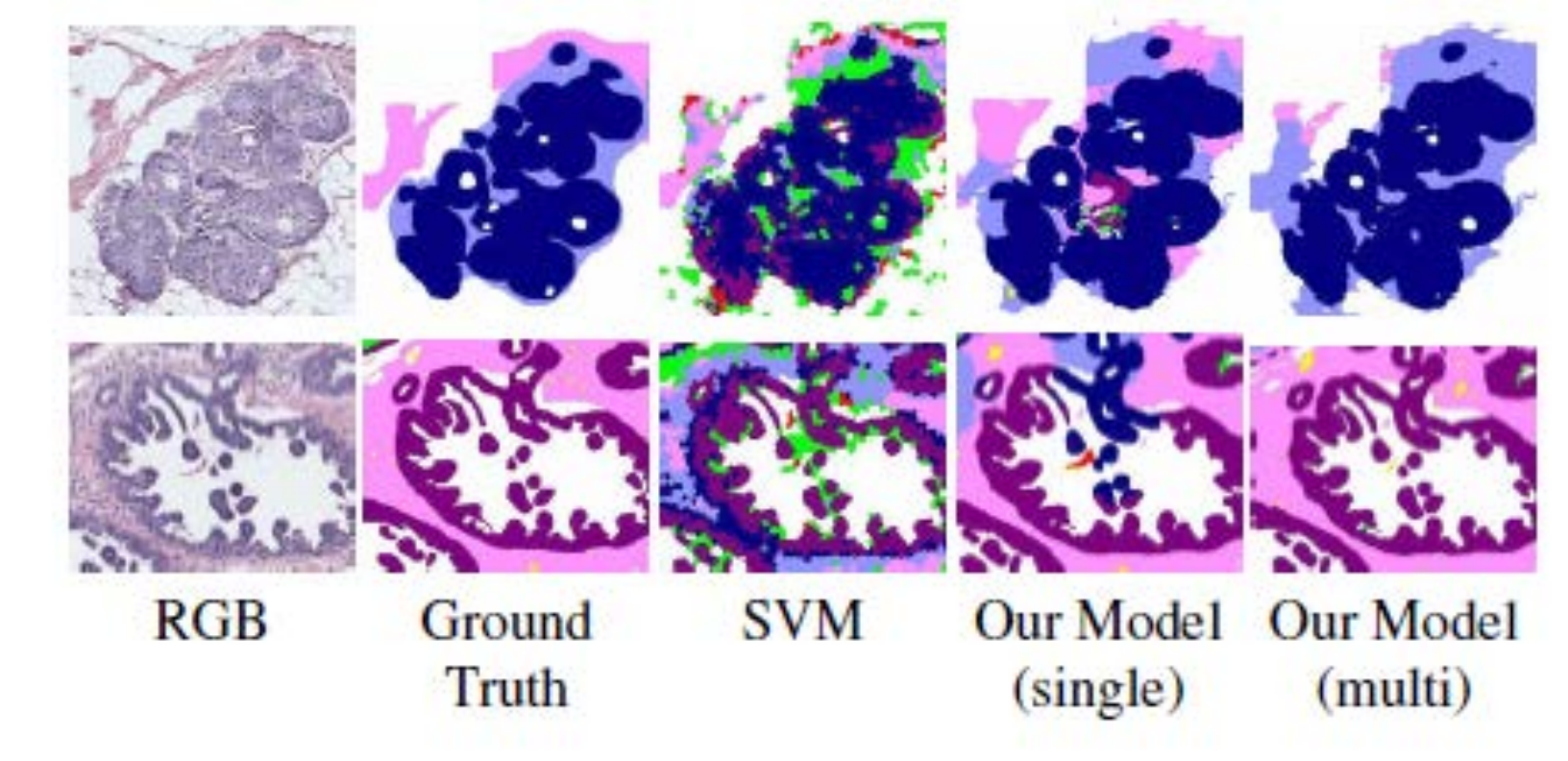}}
\caption{Segmentation result. The figure corresponds to Fig.9 in~\cite{Mehta-2018-LSBB}.}
\label{fig:LSBBR}
\end{figure}

In~\cite{Mehta-2018-LSBB}, a new encoder-decoder architecture is proposed to deal with the semantic segmentation problem of breast biopsy WSI. The designed new architecture includes four new functions: (1) Input Perceptual Coding Block (IA-RCU), which can enhance the internal input of the encoder to compensate for information loss caused by downsampling, (2) densely connected decoding network, and (3) additional sparse connection of the decoding network to effectively combine the multi-level features of the encoder aggregation, and (4) a multi-resolution network for context-aware learning, which uses densely connected fusion blocks to combine rate outputs of different resolutions. The result after segmentation is shown in Figure.~\ref{fig:LSBBR}.

\begin{figure}[htbp!]
\centerline{\includegraphics[width=0.98\linewidth]{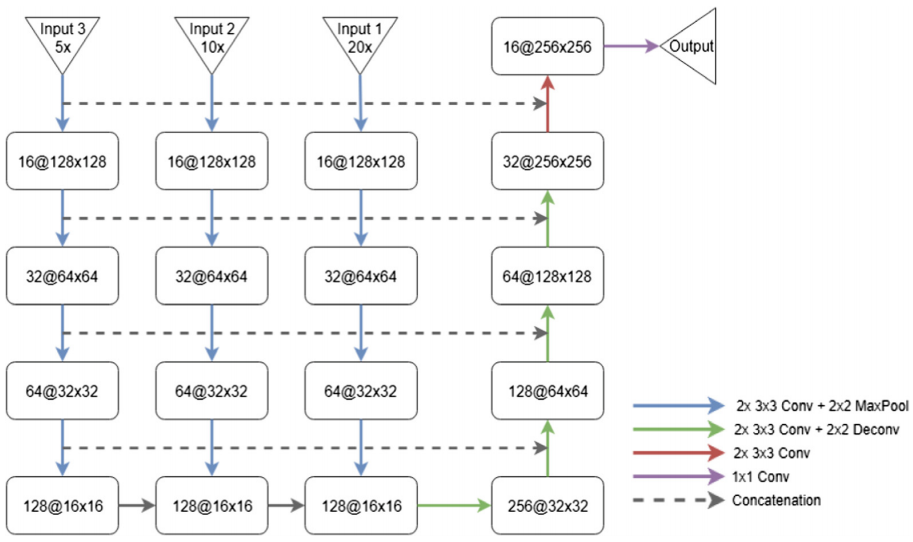}}
\caption{U-net structure used in~\cite{Seth-2019-ASDW}. The figure corresponds to Fig.2 in~\cite{Seth-2019-ASDW}.}
\label{fig:ASDW}
\end{figure}

In~\cite{Seth-2019-ALBD} and~\cite{Seth-2019-ASDW}, several U-net architectures and DCNN, which are used to output probability maps, are trained to segment the WSI of ductal carcinoma in situ (DCIS) in wireless sensor networks, and the minimum required level of a good patch vision for achieving excellent accuracy on slides is verified. U-net is trained five times to achieve the best test results ($DSC = 0.771$, $F1 = 0.601$), which means that U-net can benefit from broader context information. The U-net structure used is shown in the Figure.~\ref{fig:ASDW}.

In~\cite{Feng-2020-DLMA}, a multi-scale image processing method is proposed to solve the image segmentation problem, that is, the localization problem in the histopathological imaging of liver cancer. These eight networks are compared, and the most suitable network for liver cancer segmentation is obtained. Through comprehensive performance comparison, U-net is selected. The local color normalization method of the pathological image is used to solve the influence of the background, and then a seven-layer Gaussian pyramid is used to represent it. Multi-scale image sets are then created for each WSI. Well-trained U-net, fine-tuned at each level to obtain independent models. Then, shift cropping uses weighted overlap in the prediction process to solve the problem of block continuity. Finally, the predicted image is mapped back to the original size, and voting is performed, and a multi-scale predicted image fusion mechanism is proposed. The experimental data is the verification image of the 2019 MICCAI PAIP Challenge. This evaluation result shows that the algorithm is better than other algorithms.

Table.~\ref{SCAD} summarizes the work of different teams using ANNs to analyze WSI for segmentation tasks.
\begin{table*}[htbp!]\tiny
\centering
\caption{Summary of WSI segmentation tasks by different teams using artificial neural network (ANN).}
\label{SCAD}
\renewcommand\arraystretch{1.5}
\newcommand{\tabincell}[2]{\begin{tabular}{@{}#1@{}}#2\end{tabular}}
\setlength{\tabcolsep}{10mm}
\begin{tabular}{@{}ccccc@{}}
\toprule
Year & Reference                  & Team                                            & Method                                                                                        & Evaluation Index                                                     \\ \hline
2017 & ~\cite{Bandi-2017-CDMT}        & Bándi P                                       & FCNN, U-Net                                                                                    & Jaccard index                                                        \\
2017 & ~\cite{Xu-2017-LSTH}               & Xu Y                                            & SVM-CNN                                                                                        & Accuracy     \\
2018 & ~\cite{Cui-2018-DLAO}            & Cui Y                                           & FCN                                                                                           & \tabincell{c}{precision, recall, F1,\\ Dice co-efficient}                               \\
2018 &~\cite{Sirinukunwattana-2018-IWSS} &  K Sirinukunwattana                             & CNN                                                                                           & \tabincell{c}{Classification accuracy as\\ measured by the F1-measure}                \\
2018 & ~\cite{Mehta-2018-LSBB}            & Sachin Mehta                                    & \tabincell{c}{multi-resolution \\encoder-decoder network}                                                      & F1, mIOU, PA                                                           \\
2018 & ~\cite{Dong-2018-RANT}             & Nanqing Dong                                    & \tabincell{c}{Reinforced Auto-Zoom Net\\(RAZN)}                                                                & \tabincell{c}{intersection over union (IOU),\\ Relative inference time}               \\
2019 & ~\cite{Seth-2019-ALBD}             & Nikhil Seth                                     & \tabincell{c}{original U-Net architecture\\ to use ELU and\\ a multi-resolution network, \\Ensemble models}         & \tabincell{c}{Dice Score, Dice P-Value, \\F1 Score, F1 P-Value, Time(s)}           \\
2019 & ~\cite{Seth-2019-ASDW}             & Seth N, Akbar S                                 & U-Net                                                                                         & Dice co-efficient, Modified F1 Score                                   \\
2020 & ~\cite{Feng-2020-DLMA}             & Feng Y, Hafiane A                               & U-net    & \tabincell{c}{F1 score, Jaccard similarity score\\ and directed Hausdorff distance} \\ \hline
\end{tabular}
\end{table*}

\subsection{Classification Methods}
\label{s:CM} 
In this section, the relevant contents of using deep learning algorithms to classify histopathological WSIs are summarized.

Metastatic breast cancer needs to be identified in~\cite{Wang-2016-DLIM}. Firstly, the background is detected automatically based on the threshold. By comparing the performance of GoogLeNet, AlexNet, VGG16 and FaceNet, GoogLeNet is selected to generate tumor probability thermogram and RF classifier is used to classify metastatic WSI and negative WSI, and finally AUC of $0.9254$ is obtained.

In~\cite{Sheikhzadeh-2016-ALMB}, the complete convolution network is used to automatically label the molecular biomarkers of the complete IHC image. First, the RGB image is input into FCN, and the output is the image of different biomarkers. Each cell is classified according to the biomarkers it expressed. The results show that the proposed method is more effective than the manual marking method (the average F-score is 0.96). The neural network training process is divided into two parts: (1) Training CNN to label biomarkers expressed in cells; (2) converting the trained CNN into FCN to locate all cells expressing different biomarkers in WSI. The purpose of this paper is to train CNN through the kernel images separated from WSI and transform the trained CNN into end-to-end pixel to pixel networks like FCN, which can locate WSI biomarkers and output heat maps expressed by different biomarkers.

In~\cite{Jamaluddin-2017-TDWS}, normal sections and tumor sections are classified from histological images of lymph node tissues. The first step is to delete the useless information. Then the CNN model is reconstructed or trained to segment the tumor region. The details are in
Sec.~\ref{ss:int:DM}. After the tumor region is segmented, features are extracted and RF classifier is used for classification. Finally, the AUC score is $0.94$.

\begin{figure}[htbp!]
\centerline{\includegraphics[width=0.95\linewidth]{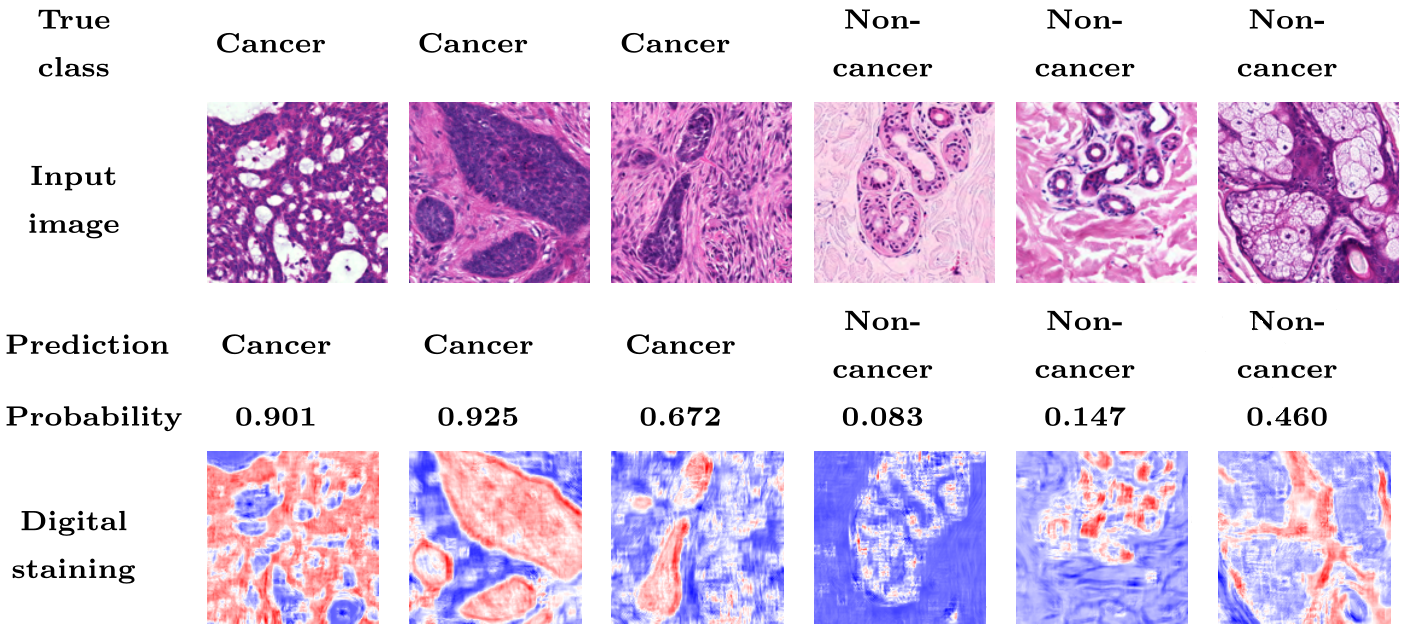}}
\caption{Outputs produced by the system for different cancer and non-cancer input images. The figure matches to Table.2 in~\cite{Arevalo-2015-UFLF}.}
\label{fig:UFLF}
\end{figure}

In~\cite{Arevalo-2015-UFLF}, WSIs of basal cell carcinoma are studied as a complete unsupervised feature. First, there is a set of feature detectors. This feature detector is selected from a set of patches randomly selected from the collection of images. Then this group of feature detectors is used to capture the most common patterns, and the method used is to simulate an auto-encoder neural network. Then convolution or BOF method is used to represent the image. This representation is implemented using the feature detector learned in the previous step. Next, the binary classification model, the softmax regression classifier, is trained from the representation obtained by the convolution or BOF method. Basal cell carcinoma includes different types of cancer and non-cancer. The final result of the system is shown in Figure.~\ref{fig:UFLF}. 
The best result is obtained in AUC, which is better than the most advanced  $7\%$ and $98.1\%$.

\begin{figure}[htbp!]
\centerline{\includegraphics[width=0.95\linewidth]{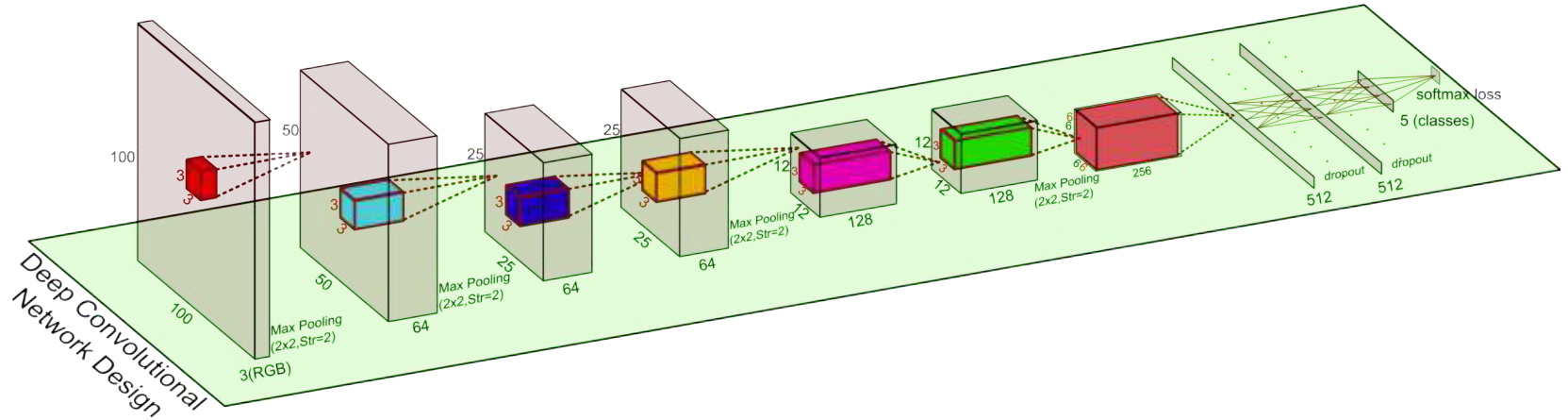}}
\caption{Designs of the CNN. 
The figure matches to Figure.4.6 in~\cite{Geccer-2016-DCBC}.}
\label{fig:DCBC}
\end{figure}

In~\cite{Geccer-2016-DCBC} and~\cite{Gecer-2018-DCCW}, a deep convolution network is used to detect and classify WSI and ROI in breast cancer. The detection part is described in detail in Sec.~\ref{ss:int:DM}. This part mainly introduces the classification, that is, ROI includes five diagnostic categories. A classifier is designed based on CNN, and the features obtained by CNN are used to classify the detected ROI. The structure of CNN is shown in Figure.~\ref{fig:DCBC}. Then the WSI classification is processed. WSI classifies according to the predictions of most categories in the remaining cancer areas. The results show that the efficiency is increased about 6.6 times with enough accuracy.

In~\cite{Sirinukunwattana-2016-LSDL}, it mainly includes two parts, namely the detection and classification of nuclei. The detection part proposes a space-constrained CNN for nuclear detection. The details are in Sec.~\ref{ss:int:DM}. The classification uses a combination of CNN and the new NEP. None of the proposed detection and classification methods require kernel segmentation. Then the method is applied to colorectal adenocarcinoma WSIs, and the final classification result obtained a higher F1 score than the traditional method. 

In~\cite{Hou-2016-PCNN}, an expectation-maximization method is proposed. The spatial relationship of patches is used to locate the distinguishable patches accurately. This method is suitable for the subtype classification of glioma and non-small cell lung cancer. The classification module uses CNN at patch-level and trains the decision fusion model into a two-level model. The first level (patch-level) model is based on expectation-maximization and combines the prediction of CNN output patch-level. In the second level (image-level), block-level prediction histogram is input to image-level multiple logistic regression or SVM.

In~\cite{Babaie-2017-CRDP}, a new dataset, Kimia Path24, is introduced for image classification 
and retrieval in digital pathology. The WSIs of 24 different textures are generated to test patches. 
Especially, the patch classification is based on LBP histograms, bag-of-words approach, and CNN.

In~\cite{Araujo-2017-CBCH}, a WSI classification method using CNN for breast biopsy is proposed. The proposed network architecture can retrieve information of different scales. First, the image contains 12 non-overlapping color blocks, and then the CNN and CNN + SVM classifiers trained by segments are used to calculate the probability of color block level. Finally, one of three different patch probability fusion methods is used to obtain the image classification results. The three methods are majority voting (select the most common patch as the image tag), maximum probability (select the patch category with high probability as the image tag), and probability (the category with the largest sum of patch-level probability). The final results divided the images into four categories: Invasive, in situ, benign and normal. The proposed system achieves an overall sensitivity of about $81\%$ for cancer patch classification.

In~\cite{Bejnordi-2017-CASC}, a context-aware stacked CNN Is proposed to classify WSI as normal/benign, DCIS and invasive ductal carcinoma (IDC). The first step is to train CNN and capture cell-level information with high pixel resolution. Then, the feature response generated by the model is input to the second CNN and superimposed on the first CNN. The AUC of binary classification of non-malignant and malignant slides is $96.2\%$, and the three-level accuracy of normal/benign classification of WSI, DCIS and IDC are $81.3\%$.

\begin{figure}[htbp!]
\centerline{\includegraphics[width=0.95\linewidth]{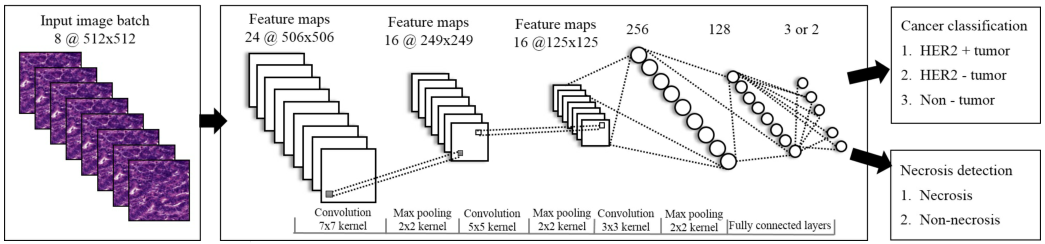}}
\caption{Proposed AlexNet architecture. 
The figure matches to Figure.4 in~\cite{Sharma-2017-DCNN}.}
\label{fig:DCNN}
\end{figure}

In~\cite{Sharma-2017-DCNN}, the automatic classification of gastric cancer WSI is proposed and compared with the traditional image analysis method and deep learning method. In the traditional analysis method, GLCM, Gabor filter, LBP histogram, gray histogram, HSV histogram and RGB histogram are used for classification and RF. In terms of deep learning methods, AlexNet is proposed as a deep convolution framework. The network structure is shown in Figure.~\ref{fig:DCNN}. According to the experiment, the overall 
classification accuracy of the cancer classification proposed by AlexNet is $0.6990$, and the 
overall classification accuracy of necrosis detection is $0.8144$. 

In~\cite{Korbar-2017-DLCC}, different types of colorectal polyps on WSIs are classified to 
help pathologists diagnose them. The architecture used is an improved version of the ResNet structure. WSI is divided into several patches and then applied to ResNet. If at least five patches on WSI are identified as such, the average confidence is $70\%$. If there is no cancer type 
in the patch, the WSI is considered normal. Finally, the accuracy is $93.0\%$. The recall is 
$88.3\%$; F1 scores is $88.8\%$.

\begin{figure}[htbp!]
\centerline{\includegraphics[width=0.95\linewidth]{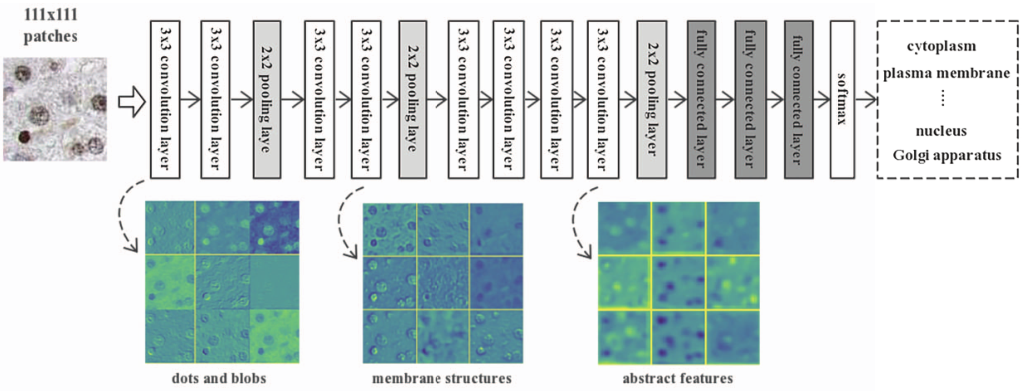}}
\caption{The network architecture diagram and the visualization of the middle layer in~\cite{Hu-2017-DLBC}.
The figure matches to Figure.4 in~\cite{Hu-2017-DLBC}.}
\label{fig:DLBC}
\end{figure}

In~\cite{Hu-2017-DLBC}, a DNN (an 11-layer CNN model) is trained to automatically learn effective features and classify protein images of 8 subcellular locations. Firstly, image preprocessing and data balance are carried out. Then, the processed data is passing through the 11 layers of the CNN model. The details of the structure are shown in Figure.~\ref{fig:DLBC}. Finally, the classification accuracy of test data is $47.31\%$, and the classification accuracy of training data is $100\%$.

In~\cite{Xu-2017-LSTH}, a deep CNN is proposed to classify, segment, and visualize large-scale histopathological images. In the part of the classification, firstly, WSI is divided into patches, then the background is discarded, and then the selected patches are input into the network to obtain a 4096-dimensional CNN feature vector. The final feature vectors are combined by softmax. Function selection is then performed to remove redundant and irrelevant functions. Finally, SVM is used for classification and the MICCAI challenge dataset is used for the experiment. The accuracy of classification is $97.5\%$.

In~\cite{Korbar-2017-LUHD}, an image analysis method based on deep learning is proposed to classify different polyp types on WSI. The classification model is based on ResNet architecture, with minor modifications. Specifically, the last fully connected layer is replaced by a convolution layer. The final result is that the average accuracy is $91.3\%$, which is better than other deep learning architectures.

\begin{figure}[htbp!]
\centerline{\includegraphics[width=0.95\linewidth]{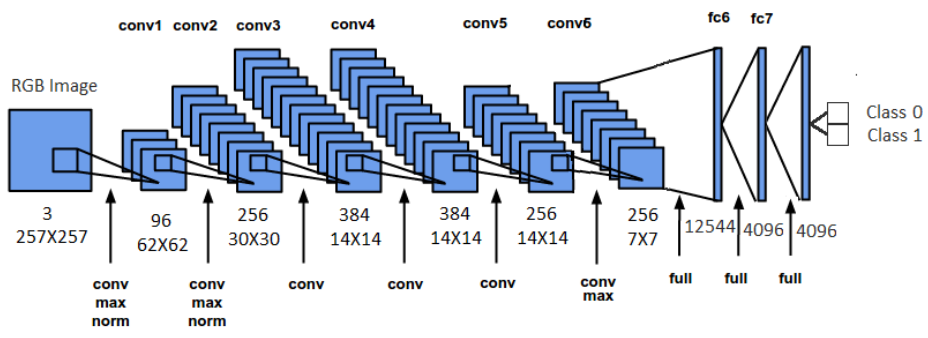}}
\caption{Architecture of the DCN used in approach. 
The figure matches to Figure.1 in~\cite{Ghosh-2017-SLCA}.}
\label{fig:SLCA}
\end{figure}

In~\cite{Ghosh-2017-SLCA}, a new deep learning method is proposed to classify white blood cells in WSI. The network uses the average pool level to find the hot spots of white blood cells in WSI. The first is to fine-tune the pre-trained AlexNet according to the dataset. Then, the network is trained on the patch dataset, and the trained network structure is shown in
Figure.~\ref{fig:SLCA}

In~\cite{Das-2017-CHWS}, a network structure based on CNN is proposed. The tissue sections of WSI are analyzed by various resolution methods. That is to say, the class a posteriori estimation of each view at a specific magnification is obtained from CNN at a specific magnification, and then the class a posteriori estimation of random multiple views at multiple magnifications is voted and filtered to provide a slide-level diagnosis. According to the experimental results, the final 
classification accuracy is $94.67 \pm 14.60\%$, sensitivity $96.00 \pm 8.94\%$, specificity 
$92.00 \pm 17.85\%$, and F-score of $96.24 \pm 5.29\%$.

In~\cite{Ren-2018-ADAC}, prostate histopathology WSI is graded. The main method is an unsupervised 
domain adaptive method. The adaptation here is achieved through confrontation training, which can 
minimize the distribution difference of feature space between two domains (annotated source domain 
and untagged target domain) with the same number of high-level classes. The loss function also uses 
Generative Adversarial Network (GAN). Besides, a Siamese architecture is developed to standardize patches in WSI. The method is then applied to the public prostate dataset for verification. The experimental results show that this method significantly improves the classification accuracy of the Gleason score.

\begin{figure}[htbp!]
\centerline{\includegraphics[width=0.7\linewidth]{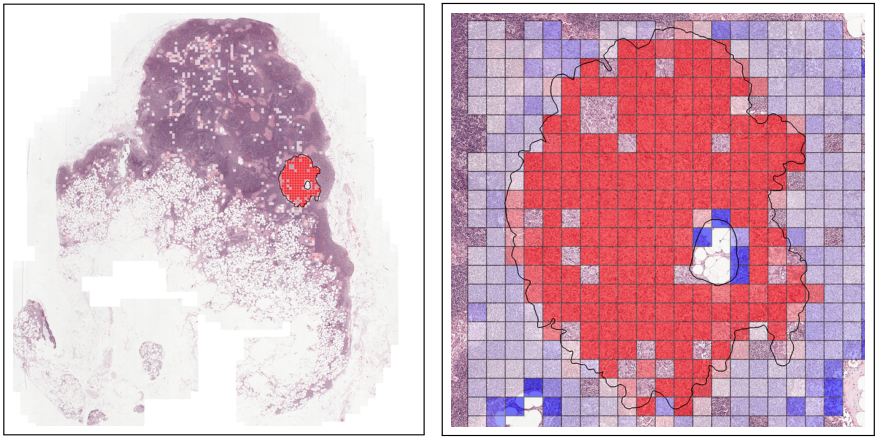}}
\caption{The classification result of the proposed method in Camelyon16 dataset. 
The figure matches to Figure.4 in~\cite{Courtiol-2018-CDLH}.}
\label{fig:CDLH}
\end{figure}

In~\cite{Courtiol-2018-CDLH}, a method of classification and localization of diseases with weak 
supervision is proposed. First, WSIs pretreatments are performed, including foreground and background 
segmentation, color normalization, and tile partitioning. Then ResNet is used for feature extraction. 
Then, the WELDON method is proposed by Durand et al. is improved and adjusted~\cite{Durand-2016-WWSL}. 
Modification of the pre-trained deep CNN model, feature embedding, and introduction of an additional 
set of full connection layers for context re-classification from instances. The final classification 
output diagram is generated as shown in Figure.~\ref{fig:CDLH}.

In~\cite{Tellez-2018-GWSI}, a two-part method is proposed to classify WSI. Firstly, the encoder is trained in an unsupervised way, the organization block on the WSI is mapped to the embedded vector, and the stack of the WSI feature graph is formed by using the sliding window. Then, the CNN classifier is trained based on WSI compact representation. There are three types of encoders trained here: A convolutional automatic encoder (CAE), a variational automatic encoder (VAE), and a new method based on contrast training. Experiments show that the new contrast encoder is better than CAE and VAE.

\begin{figure}[htbp!]
\centerline{\includegraphics[width=0.95\linewidth]{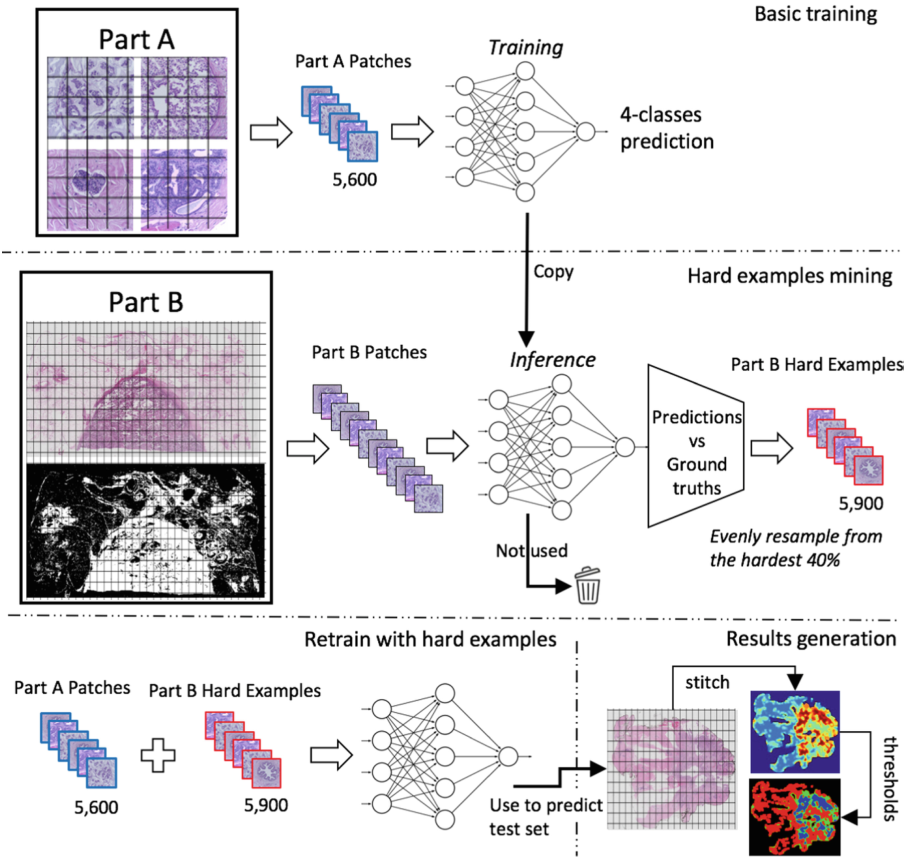}}
\caption{Overview of the framework. 
The figure matches to Figure.1 in~\cite{Kwok-2018-MCBC}.}
\label{fig:MCBC}
\end{figure}

In~\cite{Kwok-2018-MCBC}, a multi-classification of breast cancer in WSI is presented. They propose a deep learning framework, which is mainly divided into two stages, using microscopic 
images and WSI to achieve the purpose of classification. After the two types of images are patched, 
the microscopic images are used for Inception-ResNet-v2 to train the classifier. The WSIs are then 
subsampled and converted from RGB to CIE-LAB color space and segments the foreground from and background. 
Then the extraction of hard examples and patch classifier is retrained. The prediction results are 
aggregated from the block by block prediction back to image prediction and WSI annotation. This method 
is applied to ICIAR 2018 Grand Challenge on breast cancer histology images, with an accuracy rate of 
$87\%$, far exceeding second place. The specific working process is shown in Figure.~\ref{fig:MCBC}.

The author in~\cite{Das-2018-MILD} also proposes a multiple instance learning (MIL) framework based on breast 
cancer WSI. The framework is based on CNN and introduces a new pooling layer, which enables patches in WSI to aggregate most information functions. This pooling layer is a new layer of multi-instance pool (MIP), which introduces MIL into DNN as an end-to-end learning process to realize WSI classification. Finally, high classification sensitivities of $93.87\%$, $95.81\%$, 
$93.17\%$, and $88.95\%$ are achieved with four different magnifications using the public dataset.

The study in~\cite{Campanella-2018-TSDM} is also based on the MIL classification for prostate cancer. Slide tiling is the first time to run at different magnification and generate its package. Then, the model is trained to find the tile block with the highest positive and negative probability in the slide, which is used to pay more attention to the less representative examples.Based on the classification of AlexNet, ResNet, and VGG, given the threshold, if at least one instance is positive, then WSI is called positive, if all instances are negative, then the slide is negative. The optimal models are ResNet34 
and VGG11-BN, with AUC of $0.976$ and $0.977$, respectively. The method of~\cite{Campanella-2019-CGCP} 
is roughly the same as that of~\cite{Campanella-2018-TSDM}.

\cite{Yoshida-2018-AHCWG} is used to automatically evaluate the classification accuracy of gastric cancer biopsy image analysis system (E-pathologist).

\begin{figure}[htbp!]
\centerline{\includegraphics[width=0.95\linewidth]{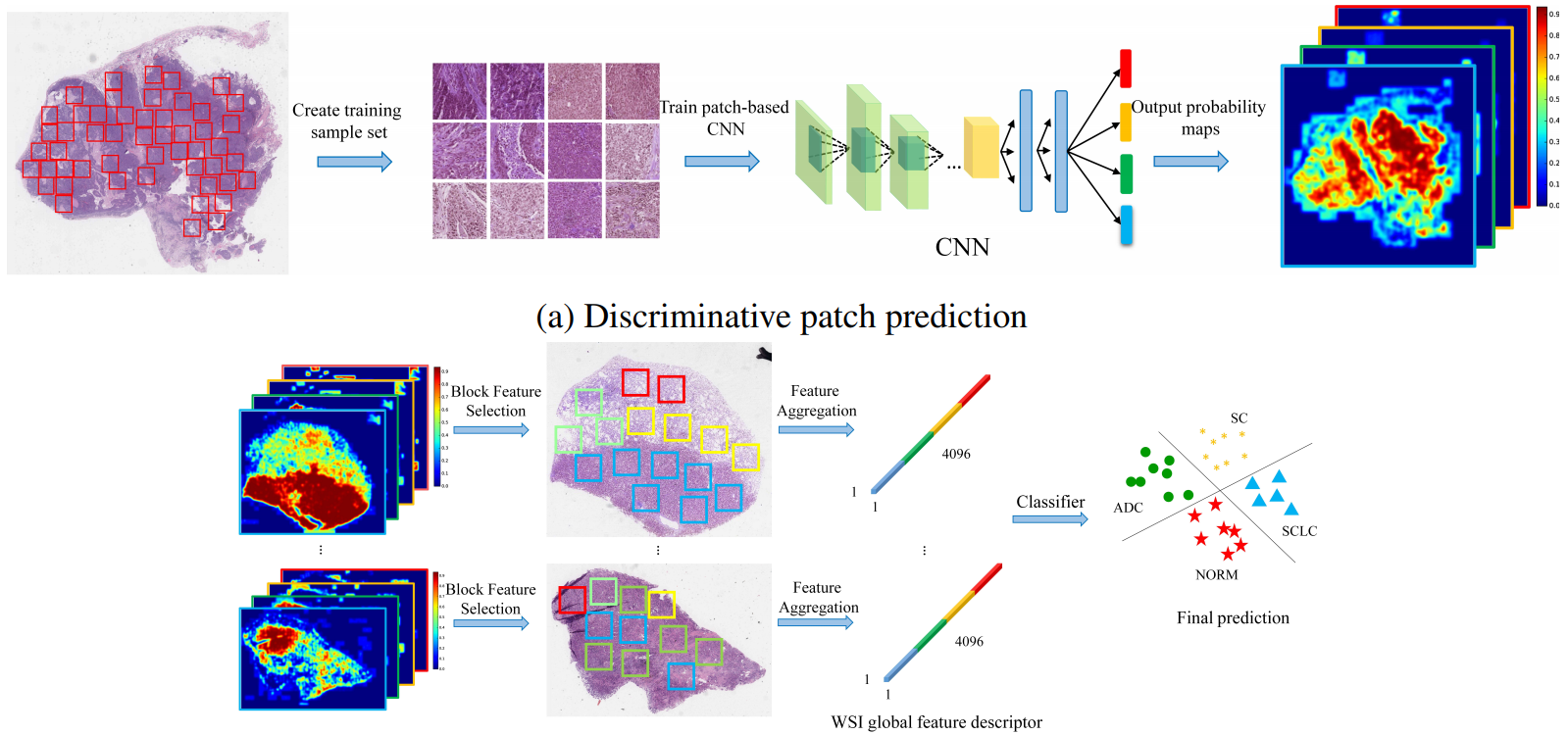}}
\caption{The whole proposed method. The figure matches to Figure.1 in~\cite{Wang-2018-WSLW}.}
\label{fig:WSLW}
\end{figure}

In~\cite{Wang-2018-WSLW}, the WSI lung cancer is classified by weakly supervised learning. Firstly, the improved FCN algorithm based on patch-level of WSI is applied to cancer prediction to find different regions as patch-level prediction models. When the probability of a block exceeds the threshold, it is retrieved. Then, context-based feature selection is performed and aggregated from the retrieved parts. The global feature descriptor is constructed. Finally, the global feature descriptor is input into the standard RF classifier. Finally, a high classification accuracy of $97.1\%$ is obtained. The whole proposed method is shown in Figure.~\ref{fig:WSLW}
.

In~\cite{Shou-2018-WSIC}, CNN is mainly used to classify gastric cancer in WSI. Firstly, the difference 
of the threshold value and color feature is used to extract the tissue and conduct morphological processing. 
Then, it is separated into patches, and the data is expanded by flipping. Then, the existing CNN 
architecture is utilized to conduct experiments on patch-level and slide-level. Finally, good results 
are obtained on DenseNet-201. And~\cite{Yoshida-2018-AHCWG} is used to automatically evaluate the classification accuracy of gastric cancer biopsy image analysis system (E-pathologist).

In~\cite{Mirschl-2018-DLCI},  the WSI of the patient is down-sampled to five times magnification, and the ROI is extracted according to the Otsu threshold and the manually refined tissue boundary. Then train the classifier to classify whether the patient has heart failure. Among them, two classifiers are trained. The first uses CNN, and the second uses traditional feature engineering methods based on WND-CHARM~\cite{Orlov-2008-WCMP}. It can be seen from the experimental results that CNN is better than WND-CHARM in terms of classification accuracy and sensitivity. It shows that cardiac histopathology is sufficient to accurately identify patients with clinical heart failure.

In~\cite{Li-2019-AMRM}, the prostate WSI is classified by multi-resolution based on attention mechanism. The attention-based MIL model is used to extract transient features. The whole process is divided into two parts. In the first part, attention-based clustering block selection is used to classify cancer and non-cancer. The second part is the research of cancer classification with higher resolution. Finally, the average classification accuracy is $85.11\%$, which realizes the new performance of prostate cancer classification.

In~\cite{Yue-2019-CCOP}, a machine learning algorithm for predicting the prognosis of colorectal cancer from WSI is introduced. First, data preprocessing is performed, including chroma normalization, color block extraction, and data enhancement. Then the classification process of unsupervised learning is carried out. The $K$-means algorithm first uses grouping to group patches from a single WSI, and then uses different random initial parameters for multiple clustering (two clustering methods are used here: Information density clustering and phenotype clustering). Then, CNN-based classifiers are used to train patches from different clusters. Finally, SVM is used to learn the cluster-level results. Finally, the results are used to predict the prognosis of patients.

\begin{figure}[htbp!]
\centerline{\includegraphics[width=0.95\linewidth]{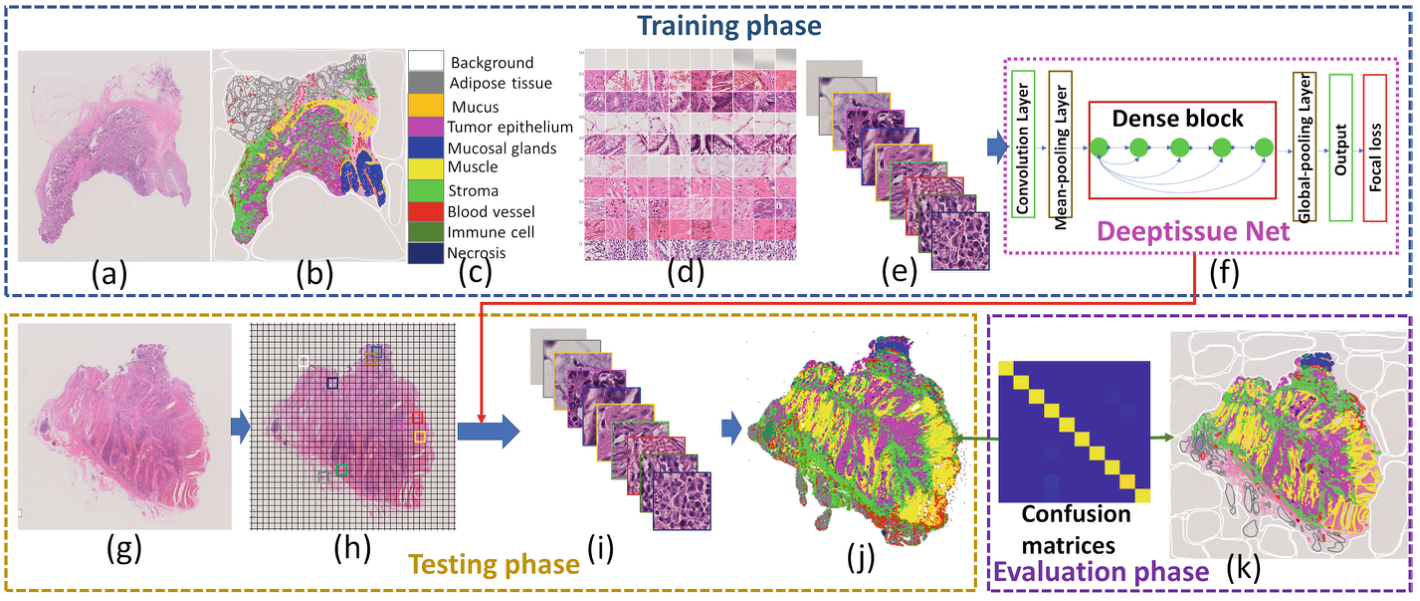}}
\caption{WSI Organization Classification Deep Organization Network Diagram. 
The figure matches to Figure.2 in~\cite{Xu-2019-MTPW}.}
\label{fig:MTPW}
\end{figure}

In~\cite{Xu-2019-MTPW}, a Deeptissue Net is used to segment WSIs of colorectal cancer histopathology images. The whole process is shown in Figure.~\ref{fig:MTPW}. (a-f) is the training process of the model. The 10 tissue sections are then marked at (a), and generate training image blocks (e) from (b) for training Deeptissue Net (f). Then test from (g) to (j) and compare the final result with manual annotation (k), and evaluate the performance through the confusion matrix. Experiments show that the classification performance of Deeptissue Net is significantly better than ResNet and DenseNet, reaching a classification accuracy of $96\%$.
.

\begin{figure}[htbp!]
\centerline{\includegraphics[width=0.95\linewidth]{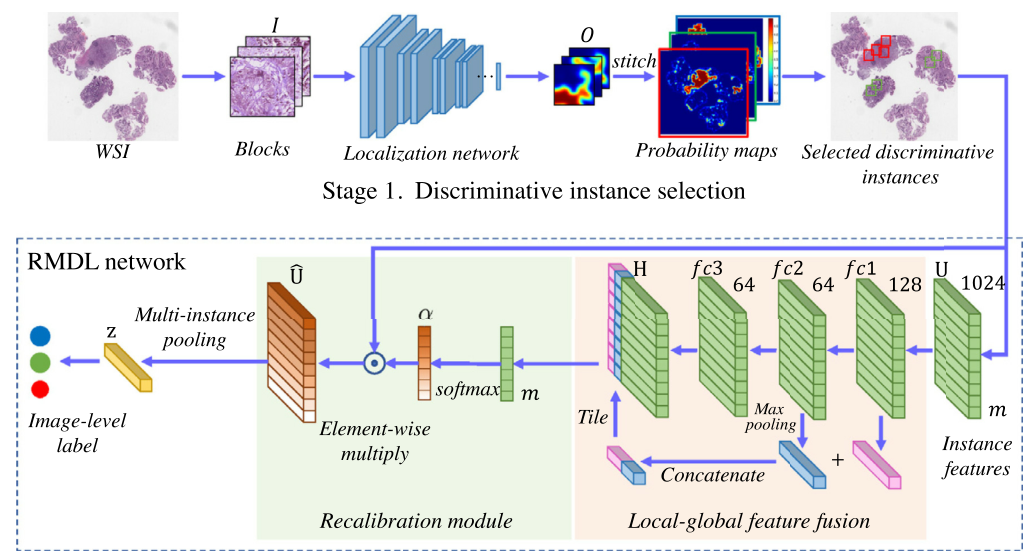}}
\caption{Overview of the framework. 
The figure matches to Figure.2 in~\cite{Wang-2019-RRMI}.}
\label{fig:RRMI}
\end{figure}

In~\cite{Wang-2019-RRMI}, multi-instance deep learning is also used to classify WSI gastric cancer images. 
A method named the RMDL method is proposed. Similar to the above MIL method, it is separated into two stages. 
In the first stage, distinguishable instances are selected, and the method is to train the localization network. In the second part, the local-global features are fused, and the RMDL network is combined by instance recalibration and a multi-instance pool module, which is used for image label prediction. The specific working process 
is shown in Figure.~\ref{fig:RRMI}.

In~\cite{Sornapudi-2020-CWSH}, a new image analysis toolbox is proposed for the diagnosis of cervical intraepithelial neoplasia (CIN) on the WSI of cervical tissue samples. The inside of the toolbox consists of four parts. The first is the preprocessing step of epithelium detection to extract the ROI. Then a binary classifier is used to filter non-epithelial regions of high-resolution interest. Then the epithelial cells are segmented using the EpithNet64 model~\cite{Sornapudi-2020-EDRE}, and preprocessed by threshold, morphology, and smoothing filter. Local cell growth pattern analysis is then performed by extracting small vertical sections. Finally, the DeepCIN model~\cite{Sornapudi-2020-FSCA} is used for CIN classification. Finally, they are integrated into an image analysis toolbox.

Table.~\ref{CCAD} summarizes the work of different teams using ANNs to analyze WSI for classification tasks.
\begin{table*}[htbp!]\tiny
\centering
\caption{Summary of WSI classification tasks by different teams using artificial neural network (ANN).}
\label{CCAD}
\renewcommand\arraystretch{1.5}
\newcommand{\tabincell}[2]{\begin{tabular}{@{}#1@{}}#2\end{tabular}}
\setlength{\tabcolsep}{10mm}
\begin{tabular}{@{}ccccc@{}}
\toprule
Year & Reference                  & Team                                                                         & Method                                                                                     & Evaluation Index                                                                                              \\ \hline
2014 & ~\cite{Cruz-2014-ADID}             & Angel Cruz-Roa                                                               & CNN, Random Forest Classifier                                                               & \tabincell{c}{TP, FP, TN, FN, Precision,\\ Recall, Sensitivity, \\Specificity, F1, BAC}                                               \\
2015 & ~\cite{Arevalo-2015-UFLF}          & John Arevalo Angel Cruz-Roa                                                  & $K$-means                                                                                  & AUC                                                                                                           \\
2016 & ~\cite{Sharma-2016-DCNN}           & Sharma H                                                                     & CNN                                                                                        & $\backslash$                                                                                                              \\
2016 & ~\cite{Hou-2016-PCNN}              & Hou L                                                                        & CNN                                                                                        & $\backslash$                                                                                                              \\
2016 & ~\cite{Wang-2016-DLIM}             & Dayong Wang Aditya Khosla                                                    & \tabincell{c}{a random forest classifier,\\ GoogLeNet}                                                      & AUC, FROC                                                                                                      \\
2016 & ~\cite{Sirinukunwattana-2016-LSDL} & Sirinukunwattana K                                                           & \tabincell{c}{Neighboring Ensemble Predictor \\(NEP) coupled with CNN}                                      & Weighted Average F1 score                                                                                     \\
2016 & ~\cite{Sheikhzadeh-2016-ALMB}      & Fahime Sheikhzadeh                                                           & FCN                                                                                                         & Accuracy and average F-score                                                              \\
2016 & ~\cite{Geccer-2016-DCBC}           & Baris Gecer                                                                  & FCN, CNN                                                                                                    & Accuracy                                                                                  \\
2017 & ~\cite{Araujo-2017-CBCH}           & Araújo T                                                                     & \tabincell{c}{patch-wise trained CNN \\and CNN+SVM classifiers}                                             & Sensitivity and Accuracy                                                                                      \\
2017 & ~\cite{Bejnordi-2017-CASC}         & Bejnordi B E                                                                 & context-aware stacked CNN                                                                  & AUC, Accuracy                                                                                               \\
2017 & ~\cite{Korbar-2017-DLCC}           & Korbar B                                                                     & a modified version of a ResNet                                                             & \tabincell{c}{Accuracy, precision, recall,\\ and F1 score and their\\ 95\% confidence intervals}                                \\
2017 & ~\cite{Hu-2017-DLBC}               & Hu J X                                                                       & CNN                                                                                        & Accuracy                                                                                                      \\
2017 & ~\cite{Das-2017-CHWS}              & Das K, Karri S P K                                                           & \tabincell{c}{CNN, majority voting over\\ random multi-views \\at multi-magnification}                         & $\backslash$                                                                                    \\
2017 & ~\cite{Sharma-2017-DCNN}           & Sharma H                                                                     & self-designed CNN architecture                                                             & Average classification accuracy                                                                               \\
2017 & ~\cite{Babaie-2017-CRDP}           & Babaie M                                                                     & dictionary approach and CNNs                                                               & Accuracy                                                                                                      \\
2017 & ~\cite{Xu-2017-LSTH}               & Xu Y                                                                         & SVM-CNN                                                                                    & Accuracy                                                                                                      \\
2017 & ~\cite{Ghosh-2017-SLCA}            & Ghosh A                                                                      & deep convolutional network                                                                 & \tabincell{c}{TP, TN, FP, FN, Acc,\\ Specificity, Sensitivity, F-score}                                                               \\
2017 & ~\cite{Jamaluddin-2017-TDWS}       & Jamaluddin M F                                                               & CNN, Random Forest                                                                         & AUC                                                                                                           \\
2017 & ~\cite{Korbar-2017-LUHD}           & Korbar B                                                                     & ResNet                                                                                     & Average IOU                                                                                                   \\
2018 & ~\cite{Ren-2018-ADAC}              & Jian Ren                                                                     & A Siamese network                                                                          & $\backslash$                                                                                          \\
2018 & ~\cite{Yoshida-2018-AHCWG}         & Hiroshi Yoshida                                                              & \tabincell{c}{used multi-instance learning\\ (MIL) for training a  multilayer \\neural network (MLNN) model} & \tabincell{c}{Sensitivity, Specificity, \\Positive predictive  value, \\Negative predictive value}                                  \\
2018 & ~\cite{Tellez-2018-GWSI}           & David Tellez                                                                 & a CNN-based classifier                                                                     & Area under the ROC curve                                                                                      \\
2018 & ~\cite{Das-2018-MILD}              & Kausik Das                                                                   & a MIL framework for CNN                                                                    & $\backslash$                                                                                                              \\
2018 & ~\cite{Wang-2018-WSLW}             & X Wang                                                                       & \tabincell{c}{FCN, ScanNet, \\Random Forest (RF) classifier}                                                & $\backslash$                                                                                                              \\
2018 & ~\cite{Shou-2018-WSIC}             & Junni Shou                                                                   & \tabincell{c}{CNN(DenseNet-201 is the\\ best structure), RF Classifier}                                      & Accuracy, Sensitivity, Specificity                                                                              \\
2018 & ~\cite{Campanella-2018-TSDM}       & Campanella G                                                                 & VGG11-BN and ResNet34                                                                      & \tabincell{c}{Accuracy, Confusion matrix,\\ and ROC curve}                                                                      \\
2018 & ~\cite{Courtiol-2018-CDLH}         & Courtiol P                                                                   & MLP Classifier                                                                             & AUC                                                                                                           \\
2018 & ~\cite{Gecer-2018-DCCW}            & Baris Gecer                                                                  & FCN, CNN                                                                                    & $\backslash$                                                                                                              \\
2018 & ~\cite{Kwok-2018-MCBC}             & Scotty Kwok                                                                  & Inception-Resnet-v2                                                                        & $\backslash$                                                                                            \\
2018 & ~\cite{Mirschl-2018-DLCI}          & Jeffrey J. Nirschl                                                           & CNN                                                                                        & Sensitivity, specificity                                                                                \\
2019 & ~\cite{Campanella-2019-CGCP}       & Gabriele Campanella                                                          & CNN and MIL                                                                                & AUC                                                                                                           \\
2019 & ~\cite{Bilaloglu-2019-EPCW}        & S Bilaloglu                                                                  & PathCNN                                                                                    & AUC                                                                                                           \\
2019 & ~\cite{Wang-2019-RRMI}             & Shujun Wang                                                                  & \tabincell{c}{a recalibrated multi-instance\\ deep learning method(RMDL)}                                   & Accuracy, Average classification score                                                                        \\
2019 & ~\cite{Yue-2019-CCOP}              & Xingzhi Yue                                                                  & $K$-means, CNN, SVM                                                                        & Accuracy, F1-score                                                                                             \\
2019 & ~\cite{Xu-2019-MTPW}               & Jun Xu                                                                       & Deeptissue Net                                                                             & Accuracy                                                                                                       \\
2019 & ~\cite{Li-2019-AMRM}               & Jiayun Li                                                                    & Vgg11bn, $K$-means clustering                                                                & Accuracy                                                                                                       \\
2020 & ~\cite{Sornapudi-2020-CWSH}        & Sudhir Sornapudi                                                             & DeepCIN model                                                                              & P, R, F1, ACC, AUC, AP, MCC             \\ \hline
\end{tabular}
\end{table*}

\subsection{Detection Methods}
This part introduces some detection methods of DNN for histopathology WSIs to improve the work efficiency of pathologists.
\label{ss:int:DM}
\subsubsection{CNN-based Deep Learning Detection Method}
The following~\cite{Zanjani-2018-CDHW,Cruz-2017-ARIB,Bejnordi-2017-DADL,Tellez-2018-WSMD,Kohlberger-2019-WSIF,Jamaluddin-2017-TDWS,Cruz-2014-ADID} are all detection classifiers based on CNN. 

In~\cite{Zanjani-2018-CDHW}, conditional random field (CRF) is applied to the potential space of deep CNN after training, and the compact features extracted from the middle layer of CNN are regarded as the observations in the fully connected CRF model to detect invasive breast cancer. The experimental results showed that the average FROC score of tumor area detection in histopathological WSI increased by about $3.9\%$. The proposed model is trained on Camelyon17 ISBI Challenge the data set and finish second with a kappa score of 0.8759.

In~\cite{Cruz-2017-ARIB}, involved images from five different cohorts from different 
institutions/ pathology labs in the United States of America and TCGA. The training dataset 
has 349 estrogens receptor-positive (ER+) invasive breast cancer patients. The approach 
yielded a Dice-coefficient of $75.86\%$, a positive predictive value of $71.62\%$ and a negative 
predictive value of $96.77\%$ in terms of pixel-by-pixel evaluation compared to manually annotated 
regions of IDC.

In~\cite{Bejnordi-2017-DADL}, the authors evaluated the performance of automatic deep learning algorithm in detecting H\&E lymph node metastasis of breast cancer women, and compared it with the diagnosis of pathologists in the diagnostic environment. This dataset is a sample of 399 breast cancer patients from RUMC and UMCU. The algorithm in this paper is superior to the artificial algorithm of the pathologist.

In~\cite{Tellez-2018-WSMD}, a method for training and evaluating CNN in WSI mitosis detection of breast cancer is proposed. In the three tasks of Tupac challenge, the performance of the proposed method is evaluated independently.

In~\cite{Kohlberger-2019-WSIF}, an autofocus quality detector called ConFocus is developed to 
detect and quantify the out-of-focus area and severity of WSIs. The ConvFocus architecture resulted in Spearman ranking co-efficients of $0.81$ and $0.94$, which are better than the results obtained by the pathologist. The expected OOF pattern is also reproduced from Z-Stack scans in two experimental scanners. In addition, this paper discusses the relationship between the accuracy of state-of-the-art metastatic breast cancer monitors and OOF. It turns out to be a negative correlation.

\begin{figure}[htbp!]
\centerline{\includegraphics[width=0.98\linewidth]{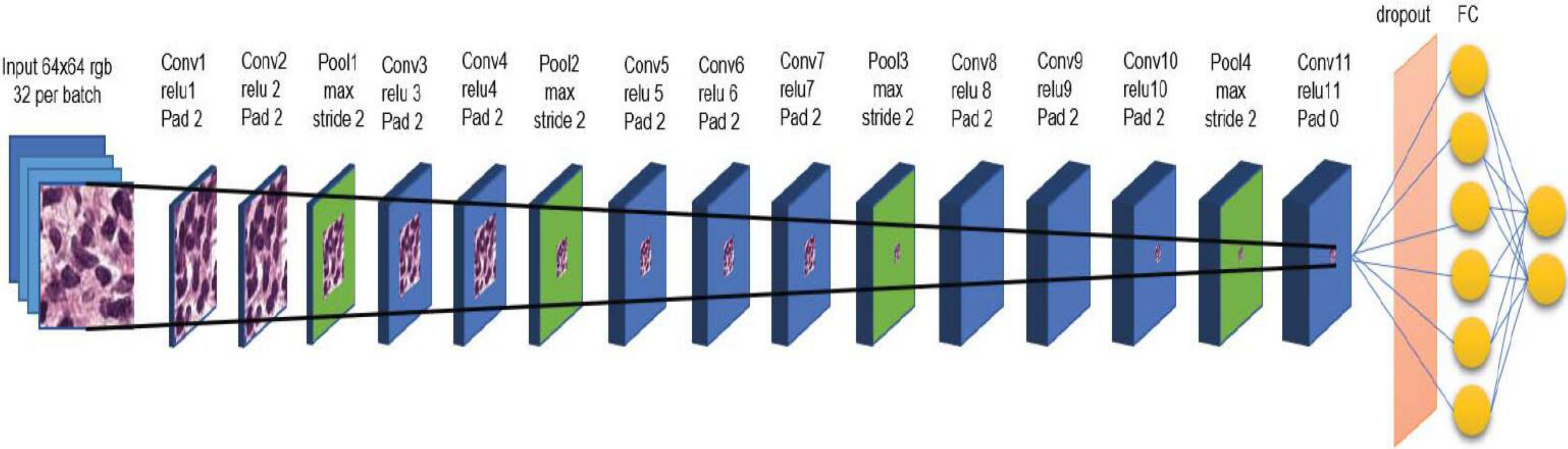}}
\caption{12 convolutional layers including the fully connected layer. 
This model was also inspired by the VGG model.}
\label{fig:fig9}
\end{figure}

\cite{Jamaluddin-2017-TDWS} is divided into two parts. The first part uses CNN to detect the possible tumor location in WSI, and the second part uses the detected results to extract features to classify normal or tumor. Using the camelyon16 dataset, which contains 160 negatives and 110 positive WSIs for training, and 50 positives and 80 negative WSIs for testing. The AUC of this method at $0.94$ is better than that of the winner of Camelyon16 challenge at $0.925$. The 
CNN structure designed in ~\cite{Jamaluddin-2017-TDWS} is shown in Figure.~\ref{fig:fig9}.

\begin{figure}[htbp!]
\centerline{\includegraphics[width=0.98\linewidth]{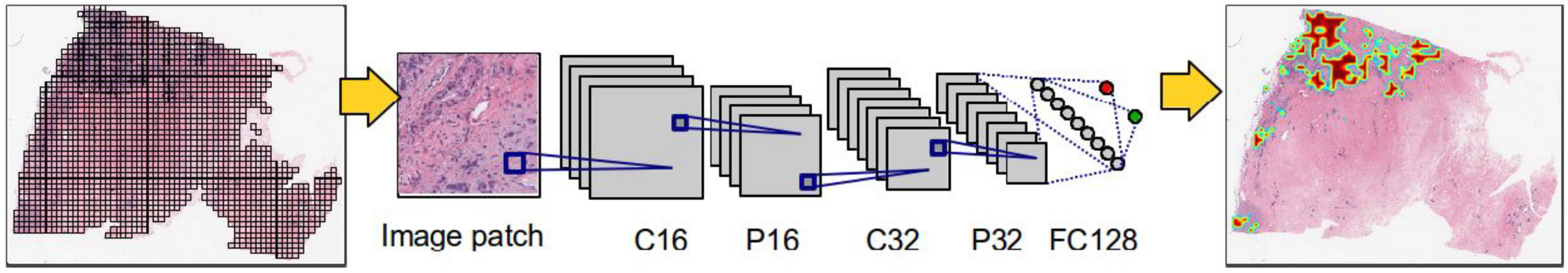}}
\caption{Overall detection framework of ~\cite{Cruz-2014-ADID}.}
\label{fig:fig10}
\end{figure}

In~\cite{Cruz-2014-ADID}, CNN is trained to detect IDC in WSI.  In the end, $71.80\%$ of 
F-measure (F1) and $84.23\%$ of balanced accuracy are obtained. The overall detection framework 
of~\cite{Cruz-2014-ADID} is shown in Figure.~\ref{fig:fig10}.

\subsubsection{FCN-based deep learning detection method}
The following~\cite{Geccer-2016-DCBC,Gecer-2018-DCCW,Lin-2018-SFDS,Lin-2019-FSFD} are processed 
with 4-layers of FCN to achieve the purpose of detection.

The study in~\cite{Geccer-2016-DCBC} includes two tasks: retrieval and classification. The retrieval part consists of four layers of FCN. Through FCN-1 feedforward processing, obvious regions are detected from WSI, and each connected component above the threshold is amplified on the input image, which is processed by FCN-2. This process lasts four times and can detect an important area of WSI. Then, CNN classifies and finally determines the diagnosis result of breast cancer.

In~\cite{Gecer-2018-DCCW}, a WSI diagnosis system for breast biopsy is proposed. First, four FCNs are used for saliency detection and multi-scale localization of ROI. Then, the convolution network is used for cancer classification. Finally, the saliency map and classification map are combined. The test results show that there is no difference between the accuracy and the prediction of pathologists.

In~\cite{Lin-2018-SFDS}, an improved FCN layer is used to input the WSI of any size, and the standard FCN layers are converted into the anchor layer. Through the anchor layer and the fast and dense ScanNet, the network can be made faster. Its cancer metastasis detection results show excellent performance on the Camelyon16 challenge dataset. A similar method is applied in~\cite{Lin-2018-SFDS} and~\cite{Lin-2019-FSFD}.

\subsubsection{Other deep learning detection methods}
The following references~\cite{Cruz-2018-HTAS,Sirinukunwattana-2016-LSDL,Bilaloglu-2019-EPCW} are used for the detection after combining CNN structure with other methods or simplifying them.

In~\cite{Cruz-2018-HTAS}, a new efficient adaptive sampling method based on probability gradient and Quasi Monte Carlo sampling is used in combination with the CNN classifier. It is suitable for the detection of invasive breast cancer on WSI. The experimental results show that Dice co-efficient is $76\%$, which is an effective strategy.
     
In~\cite{Sirinukunwattana-2016-LSDL}, in conventional colon cancer WSI proposes a space-constrained 
CNN for nuclear detection. The new NEP is used in conjunction with CNN to more accurately predict 
the type of cell detection. The test result shows that this article produces a higher average F1 score of detection.

In~\cite{Bilaloglu-2019-EPCW}, a simplified CNN, named PathCNN, is used to detect outliers of WSI 
in whole cancer. The WSI used in the experiment is downloaded from the genome data sharing database 
of the TCGA.

\subsection{Other deep learning Methods}
In this part, we select the other detection methods that appear in the papers and summarize these methods.

\begin{figure}[htbp!]
\centerline{\includegraphics[width=0.98\linewidth]{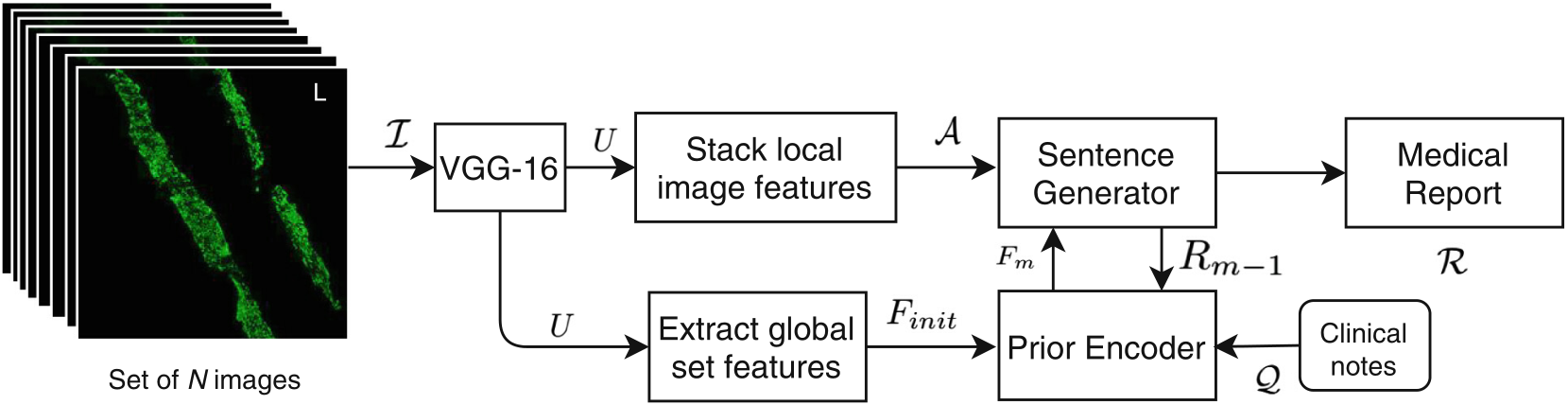}}
\caption{This image illustrates the framework of the proposed CORAL8 architecture. The figure matches to Figure.1 in~\cite{Maksoud-2019-CCOR}.}
\label{fig:Maksoud2019CCOR}
\end{figure}

In~\cite{Maksoud-2019-CCOR}, a new attention-based multi-mode RNN architecture, called CORAL8, is proposed for direct renal immunofluorescence (RDIF) detection, which solves the problem of generating medical reports for multiple image panels. Among them, the prior encoder learns to extract the context features of doctors' clinical notes, while VGG16 extracts the local and global features of WSIs. Then the two are input into RNN (sentence generator) to get the final medical report, and machine-generated text is realized. The process is shown in the Figure.~\ref{fig:Maksoud2019CCOR}.

\begin{figure}[htbp!]
\centerline{\includegraphics[width=0.98\linewidth]{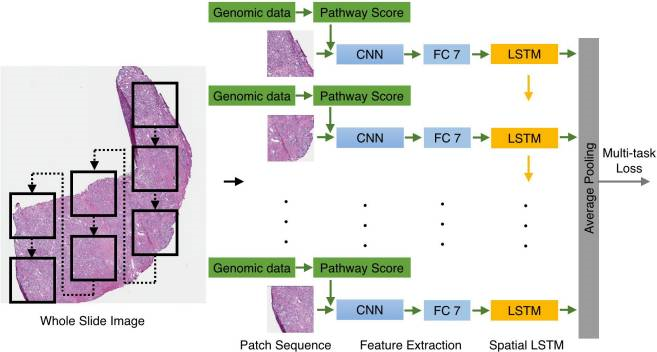}}
\caption{The architecture of convolutional neural networks (CNN) with LSTM. The figure matches to Figure.1 in~\cite{Ren-2018-DPCP}.}
\label{fig:Ren2018DPCP}
\end{figure}

In~\cite{Ren-2018-DPCP}, the CNN architecture of LSTM is used to analyze the risk of disease progression and recurrence in prostate cancer patients with the Gleason score of 7. The data used in this study came from TCGA, which identified the image biomarkers of WSI and divided the Gleason score of prostate cancer patients into $3 + 4$ or $4 + 3$. The results show that the combination of image features and genomic pathway score is more suitable than standard clinical markers and image texture features for evaluating the possibility of recurrence. Patients with the Gleason score of $4 + 3$ has a higher risk of progression and recurrence than patients with Gleason score of $3 + 4$. The structure of the LSTM CNN using WSI and genomic data to quantify images and genetic biomarkers is shown in Figure.~\ref{fig:Ren2018DPCP}.
\label{ss:int:OM}

Table.~\ref{DCAD} summarizes the work of different teams using ANNs to analyze WSI for detection tasks.
\begin{table*}[htbp!]\tiny
\centering
\caption{Summary of WSI detection tasks by different teams using ANN.}
\label{DCAD}
\renewcommand\arraystretch{1.85}
\newcommand{\tabincell}[2]{\begin{tabular}{@{}#1@{}}#2\end{tabular}}
\setlength{\tabcolsep}{7mm}
\begin{tabular}{@{}ccccc@{}}
\toprule
Year & Reference            & Team                                                   & Method                                                                                                                                          & Evaluation Index                                                                                              \\ \hline
2014 & ~\cite{Cruz-2014-ADID}       & Angel Cruz-Roa                                         & CNN                                                                                                                                             & F-measure (F1), balanced accuracy                               \\
2016 & ~\cite{Sirinukunwattana-2016-LSDL} & Korsuk Sirinukunwattana                          & SC-CNN and NEP                                                                                                                                  & F1-score                                                         \\
2016 & ~\cite{Geccer-2016-DCBC}     & Geçer B                                                & CNN                                                                                                                                             & ROC, accuracy                                                    \\
2017 & ~\cite{Jamaluddin-2017-TDWS} & M F Jamaluddin                                         & CNN                                                                                                                                             & AUC                                                            \\
2017 & ~\cite{Cruz-2017-ARIB}       & Cruz-Roa A                                             & ConvNet                                                                                                                                         & Dice, PPV, NPV                                                 \\
2017 & ~\cite{Bejnordi-2017-DADL}   & Bejnordi B E                                           & \tabincell{c}{DCNN, Supervised classifiers\\ (such as SVMs, RF classifiers)                                                                                   }  & AUC                                                              \\
2018 & ~\cite{Zanjani-2018-CDHW}    & Farhad Ghazvinian Zanjani                              & \tabincell{c}{apply CRFs over latent spaces\\ of a trained deep CNN}                                                                                             & FROC,kappa                                                       \\
2018 & ~\cite{Tellez-2018-WSMD}     & David Tellez                                           & CNNs                                                                                                                                            & F1-score                                                         \\
2018 & ~\cite{Mirschl-2018-DLCI}    & Jeffrey J. Nirschl                                     & \tabincell{c}{CNN classifier, 1000 tree\\ Breiman-style random \\decision forest \\(comparative approach)                                                            } & \tabincell{c}{Accuracy, Sensitivity, Specificity, \\Positive predictive value, AUC} \\
2018 & ~\cite{Cruz-2018-HTAS}       & Angel Cruz-Roa                                         & \tabincell{c}{High-throughput Adaptive \\Sampling (HASHI) for \\whole-slide Histopathology Image\\ analysis, Based on CNN, \\RF and SVM}  & \tabincell{c}{Dice, PPV, NPV, TPR,\\ TNR, FPR, FNR                                     } \\
2018 & ~\cite{Gecer-2018-DCCW}      & Baris Gecer                                            & FCN                                                                                                                                             & Accuracy                                                          \\
2018 & ~\cite{Lin-2018-SFDS}        & Huangjing Lin                                          & \tabincell{c}{fast and dense scanning framework, \\referred as ScanNet (modified FCN) } & FROC, AUC                                                        \\
2019 & ~\cite{Kohlberger-2019-WSIF} & Kohlberger T                                           & ConvFocus                                                                                                                                       & Spearman rank co-efficients                                       \\
2019 & ~\cite{Bilaloglu-2019-EPCW}  & Seda Bilaloglu                                         & PathCNN                                                                                                                                         & AUC and $95\%$ confidence interval                                 \\
2019 & ~\cite{Lin-2019-FSFD}        & Huangjing Lin                                          & Fast ScanNet (Based on FCN)                                                                                                                      & FROC, AUC                                                         \\ \hline
\end{tabular}
\end{table*}
\section{Methodology Analysis}
\label{MA}
This section analyzes prominent methods in different tasks.

\begin{figure}[htbp!]
\centerline{\includegraphics[width=0.98\linewidth]{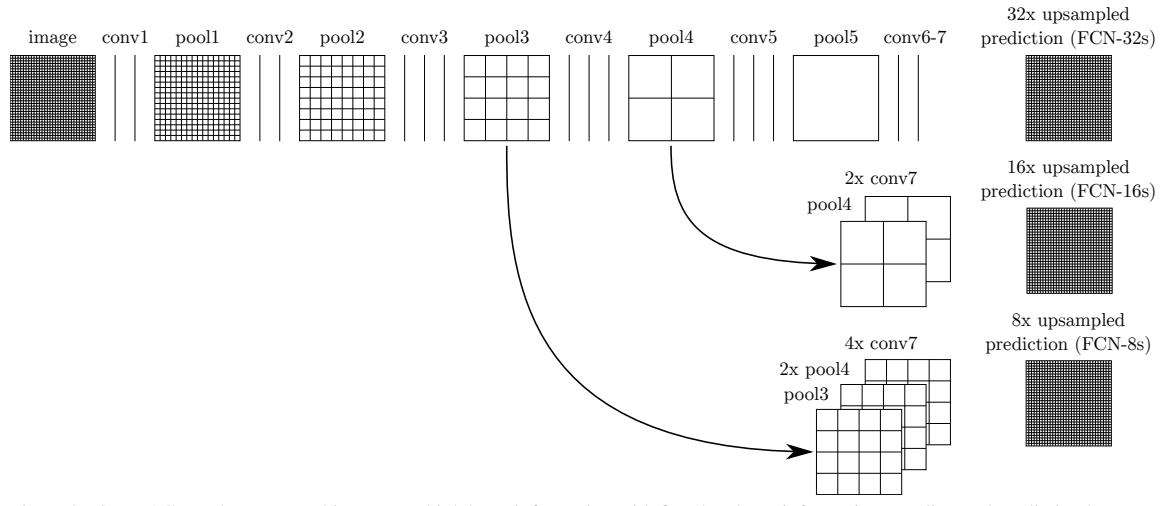}}
\caption{Jump connection structure of FCN. The figure matches to Figure.3 in~\cite{Long-2015-FCNS}.}
\label{fig:skip}
\end{figure}

We can see from the above Table.~\ref{SCAD} that FCN and U-net are mostly used in WSI segmentation using ANN. From this, we can know that the model with skip connection and encoder-decoder network like U-net is a successful model for medical images. U-net is just the extension of FCN. FCN is that there is no fully connected layer in the network, and it is replaced by the convolutional layer. In addition, the skip connection~\cite{Long-2015-FCNS} structure is used to refine the segmentation. The schematic diagram of skip connection is shown in Figure.~\ref{fig:skip}. In this way, the network used for classification can be rewritten as the pixel classification network used for segmentation. However, due to the lack of consideration of context information between pixels, the segmentation result is not precise enough. So some researchers use FCN and CRF~\cite{Sun-2020-GHIS} to further optimize the segmentation results. U-net captures both the contracting path of context information and the symmetric expanding path that allows accurate positioning, which enables the network to propagate context information to a higher resolution layer. Because the network can combine the low-resolution information after downsampling with the high-resolution information after upsampling, it is perfect for medical images with fuzzy boundaries and accurate segmentation. However, the U-net structure can only be predicted on a single scale, and can not cope with scale changes. Moreover, when the convolution layer increases, the training does not have a good generalization ability.

In the classification section, the number of cases is the largest. The classification task is similar to segmentation, but the classification focuses more on identifying and quantifying the focus area in the image.

From the Table.~\ref{CCAD}, we can see that using ANN to classify WSI, most of which use CNN-based network architecture, which can make it possible to retain all the information of the input image to the greatest extent. But there are few papers that only use deep learning algorithms. Most of them are combined with traditional machine learning algorithms, such as SVM, RF and so on~\cite{Cruz-2014-ADID,Wang-2016-DLIM,Araujo-2017-CBCH,Xu-2017-LSTH,Wang-2018-WSLW}. Some use ANN for prediction and traditional machine learning algorithms for classification~\cite{Jamaluddin-2017-TDWS}; some use ANN for pixel-level classification and traditional machine learning classifiers for slide-level classification~\cite{Shou-2018-WSIC}.

The most noteworthy thing is that starting in 2018, there has been a task of combining MIL and ANN to perform WSI classification. The concept of MIL is first used by Dietterich et al.~\cite{Dietterich-1997-SMIP} in 1997 to solve the problem of unclear training data. In MIL, the training set consists of a set of bags with classification labels. Each bag contains several instances without classification labels. If the bag contains at least one positive instance, the bag is marked as a positive bag, and vice versa. Combining MIL and ANN for classification can get better results. The CNN-based neural network is an end-to-end architecture, similar to a black box, and is weak in interpretability~\cite{Zhang-2013-CML}.

\begin{figure}[htbp!]
\centerline{\includegraphics[width=0.98\linewidth]{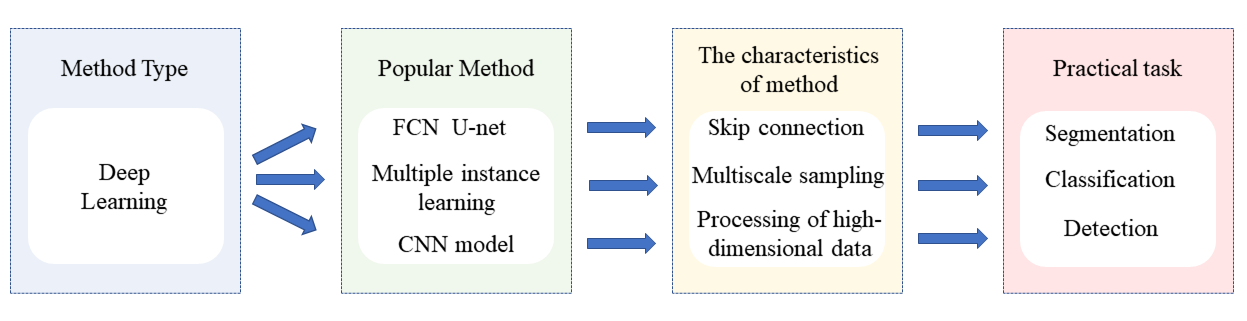}}
\caption{The popular methods of segmentation, classification, and detection.}
\label{fig:PM}
\end{figure}

The detection is generally combined with classification. Generally, WSI is tested first and then classified~\cite{Bejnordi-2017-DADL,Tellez-2018-WSMD,Lin-2019-FSFD}. The detection task generally targets the disease with breast cancer WSI~\cite{Geccer-2016-DCBC,Cruz-2017-ARIB,Zanjani-2018-CDHW,Tellez-2018-WSMD,Cruz-2018-HTAS,Lin-2018-SFDS}. Its network architecture is also based on the CNN model. The popular methods of segmentation, classification, and detection for WSIs are as shown in Figure.~\ref{fig:PM}.

\section{Potential Methods for WSI}
\label{PM}
\subsection{Attention Mechanism}
The Attention Model originally come from the study of human vision. Humans selectively focus on some parts of all information while ignoring other visible information. This mechanism is often referred to as the attentional mechanism. It was first proposed for use in natural language processing~\cite{Bahdanau-2014-NMTJ}. Attention mechanism (AM) has become an important concept in the field of neural networks. Its significance is to make the computer learn to pay attention to the key information in the image and ignore the irrelevant information~\cite{Xu-2015-SATN}. Its principle is to pay attention to the input weight allocation. It would be helpful if we could apply this mechanism in the field of WSI classification, segmentation and detection to assist doctors in diagnosis.

\begin{figure}[htbp!]
\centerline{\includegraphics[width=0.7\linewidth]{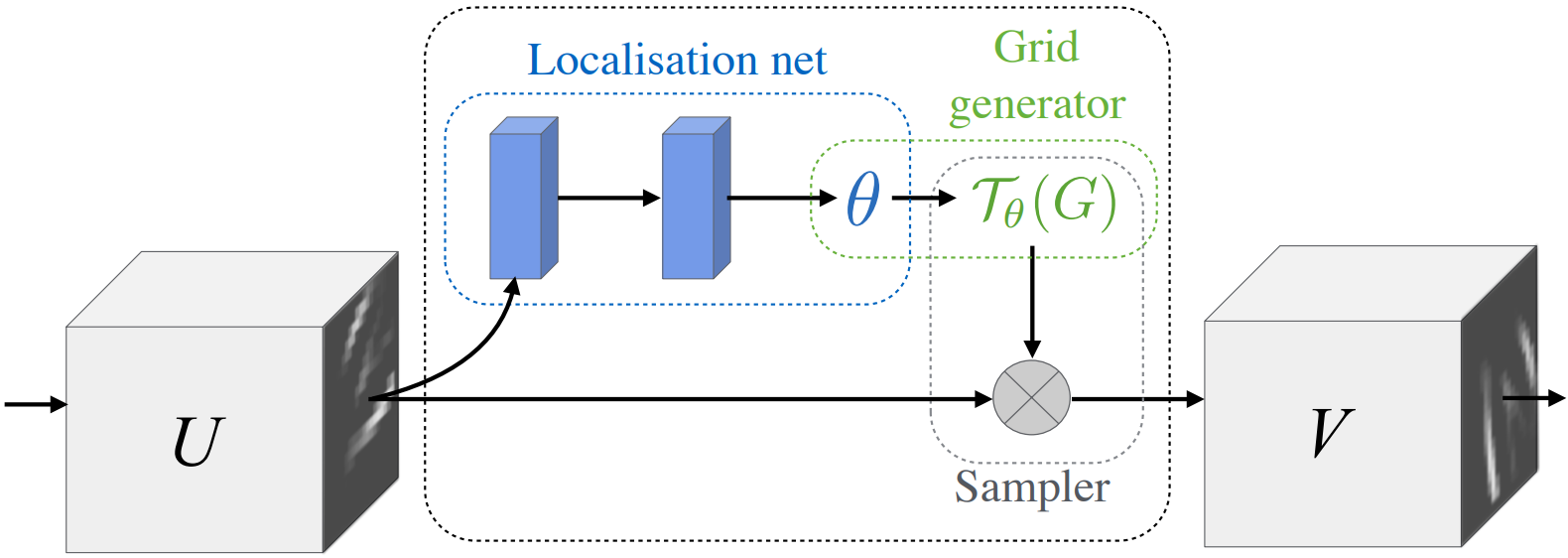}}
\caption{The architecture of a spatial transformer module. This figure matches to Figure.2 in~\cite{Jaderberg-2015-STN}.}
\label{fig:ST}
\end{figure}

Nowadays, there are many applications of AM in the field of image processing. In~\cite{Jaderberg-2015-STN}, a spatial transformer network (STN) is proposed. With this network, the ROI can be located; the ROI can be transformed into an ideal image by an affine transformation, then they are put into a neural network for training. The STN module can be put into any region of CNN as a separate module, and the input can be either the input image or the feature map. The network model used is shown in Figure.~\ref{fig:ST}.

\begin{figure}[htbp!]
\centerline{\includegraphics[width=0.98\linewidth]{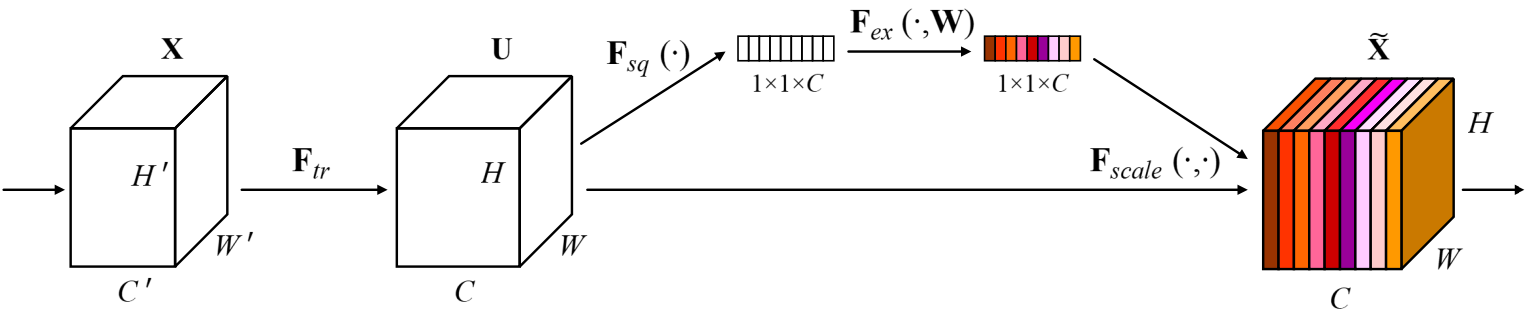}}
\caption{A Squeeze-and-Excitation block. The figure matches to Figure.1 in~\cite{Hu-2018-SEN}.}
\label{fig:SEN}
\end{figure}

In~\cite{Jaderberg-2015-STN}, AM is applied in the spatial domain, while in~\cite{Hu-2018-SEN}, the AM is applied in the channel domain. The purpose of Squeeze-and-Excitation Networks (SENET) is to learn the correlation between each channel, calculate the importance of each channel, and get a set of weight values so that the next network can selectively enhance the useful feature channel and suppress the useless feature channel. The AM module is shown in the Figure.~\ref{fig:SEN}.

In~\cite{Woo-2018-CBAM}, AM is applied in both the spatial domain and channel domain. The CBAM module proposed in this paper is divided into two parts: Channel attention and spatial attention. Channel attention is the compression of feature maps in the spatial dimension. Spatial attention is the compression of feature maps in channel dimensions. This enables the network to clearly understand which parts of the feature map should have a higher response at the channel and spatial levels. Then the module is embedded into the neural network structure for experiments.

\subsection{Visual Transformer}
The transformer is  a DNN based on the self-attention mechanism, which is initially applied in natural language processing. The emergence of the transformer solves the large-scale machine translation tasks in natural language processing. Because of the excellent performance of transformer in natural language processing, the Visual Transformer (VT) is a technology-based on the transformer in the field of computer vision~\cite{han2020survey}. The advantage of the transformer is to capture the global information by using Multi-Head Self-Attention(MHSA), to extract more powerful features. In the existing transformer-based models, some are purely transformer structure models, others are models using the combined CNN and  transformer structure. Overall, VT can better handle the classification, segmentation, and object detection tasks of pathological WSIs.\\
In~\cite{dosovitskiy2020image}, the VT model is used to process an image classification task by connecting the transformer encoder model and Multi-Layer Perceptron (MLP). It is used to cut large-scale images into the patch and position encoding for each patch to the transformer encoder model. Finally, the output features are obtained through MLP to get the final classification results. Because transformer models do not have the translation equivariance and locality of CNNs, VT is not as effective in dealing with small datasets. However, using transfer learning to pre-train VT on large-scale datasets greatly improves its accuracy, reaching an accuracy of
$88.36\%$ on ImageNet, $99.50\%$ on CIFAR-10, $94.55\%$ on CIFAR-100, and $77.16\%$ on the VTAB suite of 19 tasks. \\ 
\begin{figure}[H]
\centering
\includegraphics[scale=0.29]{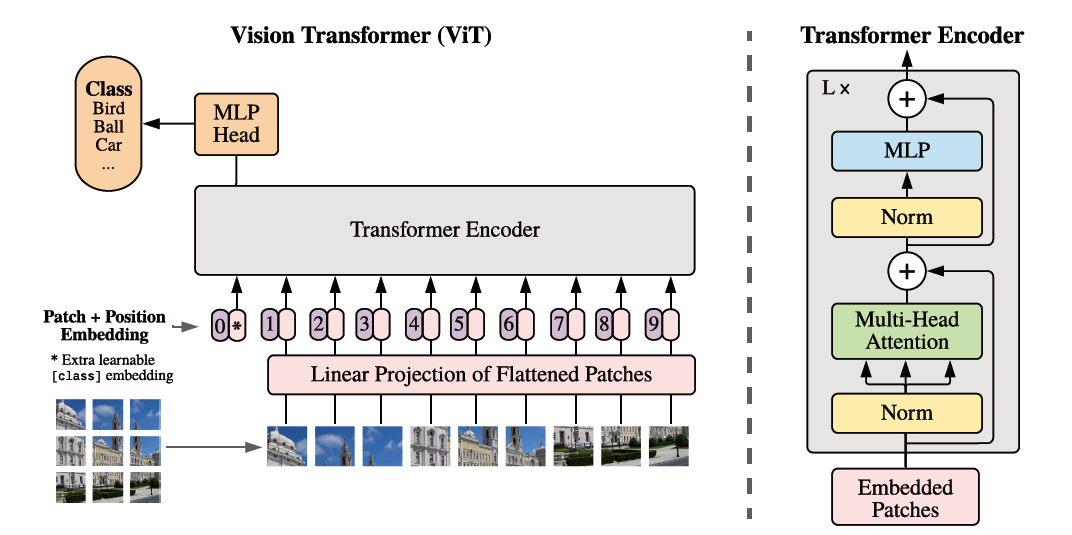}
\caption{The VT architecture.}     
\label{fig:1}
\end{figure}
In~\cite{carion2020end}, the DEtection TRansformer (DETR) model is an end-to-end object detection method by adding position encodings before the transformer. It abandons traditional hand-crafted components like anchor generation and Non-maximum Suppression (NMS) post-processing. The result is equivalent to the fast RCNN on the COCO dataset and easily migrates DETR to other tasks, such as panoramic segmentation. \\
\begin{figure}[H]
\centering
\includegraphics[scale=0.29]{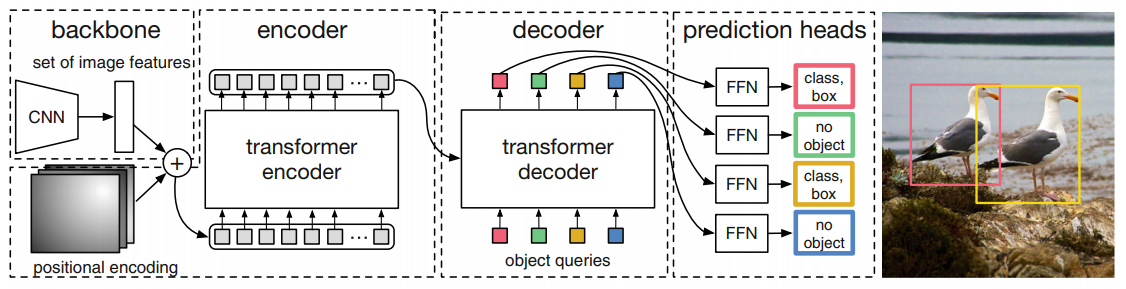}
\caption{The DETR architecture.}     
\label{fig:2}
\end{figure}
In~\cite{wang2020max}, inspired by DETR, the Max-DeepLab model directly predicts a set of non-overlapping masks and their corresponding semantic tags through a mask transformer for image segmentation. Moreover, unlike other models, which stack transformer models directly on CNNs, the Max-DeepLab model adopts a dual-path framework, which can best combine with the features extracted by CNN. \\
\subsection{Other Potential Methods Applied to WSI Technology}
In addition to AM and VT, there are some other potential methods that can be applied to WSI to facilitate its segmentation, classification, and detection.

In~\cite{Sun-2020-GHIS}, a method of gastric histopathological image segmentation based on the layered 
CRF is introduced. This method can automatically locate the cancer nest information in the stomach image, 
and because the CRF can represent the spatial relationship, a higher-order term can be established on 
this basis and applied to the image-based post-processing, which further improves the segmentation 
performance. The model shows high subdivision performance and effectiveness.

In~\cite{Galvao-2020-ISDS}, an image segmentation method is presented that is faster than superpixel is proposed. 
It separates into dense and sparse methods. Then, a new intensive method can achieve superior boundary 
adherence by exploring alternative mid-level segmentation strategies are proposed. This method is a 
very effective hierarchical segmentation method. But in this case, it applied to natural images, and 
we can also try to apply it to WSI.

In~\cite{Cui-2020-FSCA}, a method of automatic segmentation of coronary arteries based on a growing 
algorithm is proposed. Firstly, 2D U-net is used to automatically locate the initial seed points and 
the growth strategy, and then a growth algorithm combined with the 3D triangulation network is proposed. 
The improved 3D U-net is used for coronary artery segmentation. This method adopts residual block and 
two-phase training. The input data of the network is set as the neighborhood block of the seed point. 
And according to the iterative termination condition, it determines whether the segmentation is stopped.

In~\cite{Mu-2020-HICA}, a spectral space classification algorithm based on spectral space feature fusion is proposed. The algorithm is used to classify hyperspectral images by spatial coordinates. At the same time, active learning and the combination of spectrum and space features are used to improve classification performance and reduce the impact of noise.

In~\cite{Devi-2020-NPSC}, two binary classifier systems are proposed, which can classify brain magnetic resonance imaging (MRI) images. This method is novel and protects the privacy of patients. Similarly, another paper on hyperspectral image classification also deserves our attention. 

In~\cite{Park-2020-DASO}, the target in the video is detected. The method used is based on 
traditional background subtraction and artificial intelligence detection. Mask R-CNN is used to 
judge whether there is a segmentation object in the candidate region and to segment it.

This paper~\cite{Mukherjee-2020-SCAT} uses the traditional background subtraction and artificial intelligence detection to detect the target in the video. We use the mask R-CNN network architecture to judge whether there are segmented objects in the candidate region, and then segment them.

In~\cite{Mohammadi-2020-CCAG}, a feature directed network is proposed by using Multi-scale Feature Extraction Module (MFEM). The network can obtain the multi-scale context information of each abstraction level for the complex scene of the target. Moreover, a new loss function is designed, which is superior to the commonly used cross-entropy loss function. This method does not need pretreatment and can get better results. We can further apply it to histopathological WSI.

\cite{Hamadi-2020-USCM} uses two methods for target detection. The first method is to generate semantic descriptors from a set of test scores of a single concept. Then the semantic descriptor is sent to the multi-target concept detector as input. The second method is used to detect multiple concepts and categories of objects. Finally, the results of the two tests are summarized. From this paper, we can see that the combination of context semantics and features based on deep learning can produce good results.

\section{Conclusions and Future Work}
\label{s:CFW}
In this paper, a comprehensive overview of WSI analysis using ANN approaches is given, including image segmentation, classification, and detection. Within each of these tasks above, all related works are further reviewed according to their detailed categories. 

For the introduction of CAD using WSI and ANN, see Sect.~\ref{Intro}. For the commonly used ANN network architecture, see Sect.~\ref{Common}. For the commonly used WSI datasets and evaluation indicators, see Sect.~\ref{s:DEM}. In these commonly used datasets, TCGA~\cite{TCGA} and Camlyon~\cite{Litjens-2018-1HSS} are the most used two datasets. The amount of work related to WSI analysis using traditional ANNs are relatively small, and most of them are analyzed using DNNs. The reason is that since the advent of WSI technology in 1999, the large size of its image has been the reason why it is difficult for researchers to make full use of it. It was not until 2006 that related technologies began to be gradually developed and applied to pathology~\cite{Pantanowitz-2011-RCSW}. At that time, the vigorous opportunity for the development of traditional ANN has been missed. Therefore, DNNs are used to analyze histopathological WSI. See Sect.~\ref{TNNM}, Sect.~\ref{DNNM} for details. In addition to these ANN methods, we have also proposed some possible WSI methods for histopathology, such as AM and VT, see Sect.~\ref{PM} for details. In short, as more and more related technologies are proposed, they will be given to computers auxiliary diagnosis brings better progress.

In the future, the application of neural networks in WSI CAD will further improve the efficiency of pathologists, reduce the subjectivity of pathologists' diagnoses, and improve the safety of patients. Firstly, most of the diseases analyzed by WSI are focused on the breast, gastric, colon, and nervous systems. It is hoped that in the future, it can be extended to a wider range and other diseases. Secondly, there is still a lack of comprehensive, clear, fully annotated, and large-scale WSI datasets. Finally, we hope that it will be very useful to develop a network with less computation and hardware loss, higher efficiency, and interpretability.

\section*{Acknowledgements}
This work is supported by National Natural Science Foundation of China (No. 61806047). 
We thank Miss Zixian Li and Mr. Guoxian Li for their important discussion. 
We also thank M.E. Xiaoming Zhou, M.E. Jinghua Zhang and B.E. Jining Li, 
for their important technical supports.

\section*{Conflict of Interest}
The authors declare that they have no conflict of interest.

%
%

\bibliographystyle{elsarticle-num}  





\end{document}